\documentclass[acmsmall]{acmart}
\usepackage[capitalise]{cleveref}
\usepackage{subcaption}
\usepackage{algorithm}
\usepackage{algpseudocode}
\usepackage{multirow}
\usepackage{pdfpages}
\usepackage{lipsum}
\usepackage{enumitem}
\setlist[enumerate]{itemsep=0mm}
\usepackage{soul}
\usepackage{xcolor}

\usepackage{background}
\backgroundsetup{
  scale=1,
  color=black,
  opacity=0.4,
  angle=0,
  position=current page,
  vshift=12cm,
  contents={
    \parbox{0.9\textwidth}{
      \centering
      \small\textit{This article has been accepted by the
      MobileHCI 2025}
    }
  }
}

\AtBeginDocument{%
  \providecommand\BibTeX{{%
    \normalfont B\kern-0.5em{\scshape i\kern-0.25em b}\kern-0.8em\TeX}}}

\setcopyright{acmlicensed}
\copyrightyear{2024}
\acmYear{2024}
\acmDOI{XXXXXXX.XXXXXXX}

\begin{document}

\definecolor{Orange}{rgb}{0.8,0.4,0}
\definecolor{Green}{rgb}{0,0.5,0}
\definecolor{Blue}{rgb}{0,0,1}
\definecolor{Red}{rgb}{0.7,0,0}
\definecolor{Aqua}{rgb}{0,0.5,0.5}
\definecolor{airforceblue}{rgb}{0.36, 0.54, 0.66}
\definecolor{darkblue}{rgb}{0.0, 0.0, 0.55}
\definecolor{light-gray}{gray}{0.95}
\definecolor{burgundy}{RGB}{144,0,32}

\newcommand{\todo}[1]{}
\newcommand{\todofu}[1]{}
\newcommand{\todofig}[1]{\textsf{\textbf{\textcolor{Red}{[TODO FIG: #1]}}}}
\newcommand{\done}[1]{\textbf{\textcolor{Green}{[DONE: #1]}}}


\newcommand{\change}[1]{#1}
\newcommand{\changedelete}[1]{}


\newcommand{\dfkiauthor}[3]{
\author{#1}
\orcid {#3}
\email{#2}
\affiliation{%
  \institution{German Research Center for Artificial Intelligence (DFKI)}
  \city{Kaiserslautern}
  \country{Germany}
}
}

\newcommand{\dfkianduniauthor}[3]{
\author{#1}
\orcid {#3}
\email{#2}
\affiliation{%
  \institution{DFKI and RPTU}
  \city{Kaiserslautern}
  \country{Germany}
}
}


\newcommand{\ressec}[1]{\subsubsection*{$\square \square$ #1:}} 
\newcounter{takeawaycounter} \setcounter{takeawaycounter}{0}
\newcommand{\takeaway}{\subsubsection*{$\square \square$ Takeaway message \arabic{takeawaycounter}: } \stepcounter{takeawaycounter}}

\newcommand{\guideline}[1]{\subsubsection{#1}}

\newcommand{\insertfig}[5]{

\begin{figure}[t!] 
	\centering
    \includegraphics[width=#4\columnwidth]{#1}
    \caption{#3}   
    \label{#2}
    \Description{#5}
\end{figure}

}

\newcommand{\insertsubfig}[5]{
\centering
  \begin{subfigure}{#4\textwidth}
  \includegraphics[width=1.0\linewidth]{#1} 
  \caption{#3}
  \label{#2}
  \Description{#5}
  \end{subfigure}
}

\newcommand{\insertappendix}[3]{
\includepdf[scale=0.65,pages=1,pagecommand=\subsection{#1}\label{#3}]{#2}
}

\newcommand{\insertsurvey}[3]{
\includepdf[scale=0.7,pages=1,pagecommand=\section{#1}\label{#3}]{#2}
\includepdf[scale=0.8,pages=2-,pagecommand={}]{#2}
}

\newcommand{\insertdoc}[3]{
\includepdf[scale=0.8,pages=1,pagecommand=\section*{#1}\label{#3}]{#2}
\includepdf[scale=0.9,pages=2-,pagecommand={}]{#2}
}

\newcommand{\missingpax}[2]{We analyzed the data from #1 participants after excluding incomplete data from #2 participant(s).}
\newcommand{\tablestats}[3]{See Table \ref{#1} (row #2 to #3) for details about the statistics.}
\newcommand{\cone}{C1~}
\newcommand{\ctwo}{C2~}
\newcommand{\cthree}{C3~}

\newcommand{\meansd}[2]{$(Mean= #1, SD= #2)$}
\newcommand{\mean}[1]{$(M= #1)$}
\newcommand{\mrng}[1]{(means $\leq$ #1)}
\newcommand{\mrngbig}[1]{(means $\geq$ #1)}
\newcommand{\bfval}[1]{$(BF_{10}=#1)$}
\newcommand{\bfnull}{$BF_{10}$}


\title{iBreath: Usage Of Breathing Gestures as Means of Interactions}
\dfkiauthor{Mengxi Liu}{mengxi.liu@dfki.de}{0000-0003-0527-1208}
\dfkiauthor{Daniel Geißler}{daniel.geissler@dfki.de}{0000-0003-2643-4504}
\dfkiauthor{Deepika Gurung}{Deepika.gurung@dfki.de}{0009-0006-2123-828X}
\dfkiauthor{Hymalai Bello}{hymalai.bello@dfki.de}{0000-0002-9855-1040}
\dfkianduniauthor{Bo Zhou}{bo.zhou@dfki.de}{0000-0002-8976-5960}
\dfkiauthor{Sizhen Bian}{sizhen.bian@dfki.de}{0000-0001-6760-5539}
\dfkianduniauthor{Paul Lukowicz}{paul.lukowicz@dfki.de}{0000-0003-0320-6656}
\dfkiauthor{Passant Elagroudy}{passant.elagroudy@gmail.com}{0000-0003-1419-7425}

\renewcommand{\shortauthors}{Liu et al.}

\begin{abstract}
Breathing is a spontaneous but controllable body function that can be used for hands-free interaction. 
Our work introduces ``iBreath'', a novel system to detect breathing gestures similar to clicks using bio-impedance. We evaluated iBreath's accuracy and user experience using two lab studies (n=34). 
Our results show high detection accuracy (F1-scores > 95.2\%). 
Furthermore, the users found the gestures easy to use and comfortable. 
Thus, we developed eight practical guidelines for the future development of breathing gestures. 
For example, designers can train users on new gestures within just 50 seconds (five trials), and achieve robust performance with both user-dependent and user-independent models trained on data from 21 participants,  
each yielding accuracies above 90\%.
Users preferred single clicks and disliked triple clicks. The median gesture duration is 3.5-5.3 seconds. Our work provides solid ground for researchers to experiment with creating breathing gestures and interactions.
\end{abstract}



\received{06 February 2025}
\received[revised]{XX XX 2025}
\received[accepted]{XX XX 2025}

\begin{teaserfigure}
  \includegraphics[width=\textwidth]{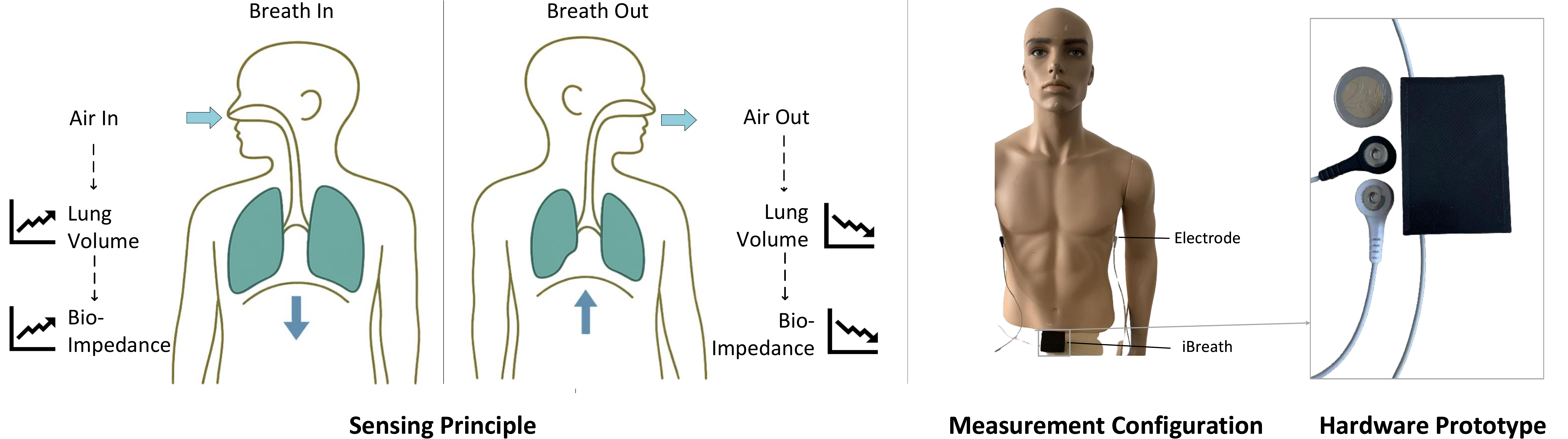}
  \caption{Sensing principle and iBreath prototype (The
core principle is that lung air volume changes upper body bio-impedance, with inhalation increasing and exhalation
decreasing it. These fluctuations are linked to variations in rhythm, speed, and intensity. We use those variations to
detect unusual breathing patterns as gestures.)}
\Description{Sensing principle and iBreath prototype (The
core principle is that lung air volume changes upper body bio-impedance, with inhalation increasing and exhalation
decreasing it. These fluctuations are linked to variations in rhythm, speed, and intensity. We use those variations to
detect unusual breathing patterns as gestures.)}
  \label{fig: sensing_principle}
\end{teaserfigure}

\maketitle

\section{Introduction}


Hands are the primary tool for most people to engage with the physical and virtual world, \change{however, hands are not always available for interaction}. Consequently, significant efforts in human-computer interaction and pervasive computing have focused on hands-free interaction, leveraging alternative body parts as actuators to support multitasking or enable interaction when hands are unavailable \cite{monteiro2021hands}. Examples range from enabling surgeons to interact with augmented reality interfaces during procedures \cite{grinshpoon2018hands} to facilitating interactions in gaming \cite{arroyo2009exploring}. Common hands-free actuators include eyes \cite{david2021towards, hakoda2017eye, ahuja2018eyespyvr, majaranta2014eye}, mouth \cite{ashbrook2016bitey, jingu2023lipio}, lips \cite{jingu2023lipio}, ears \cite{roddiger2021earrumble}, and feet \cite{minakata2019pointing, zaman2018interactive}. Beyond convenience, hands-free interaction is crucial for accessibility; for instance, 16.9\% of the elderly experience hand pain and 13.6\% suffer from hand disability \cite{dahaghin2005prevalence}, while other conditions can preclude hand use entirely \cite{vsumak2019empirical}.


Among potential hands-free modalities, breathing offers unique advantages. It is a natural, constantly available physiological process (approx. 22,000 breaths/day \cite{farhi1996breathing}) that can be consciously controlled, unlike purely autonomic signals. Controlled breathing is already practiced for applications like pain management and meditation \cite{brown2009yoga, carter2016breath}. 
 \changedelete{Controlled breathing is a subtle (almost invisible) interface, enabling discreet interaction in environments where overt body movements are restricted. Additionally, it is a hands- and eyes-free interaction. Thirdly, breathing is accessible to virtually anyone providing a valuable interaction interface to differently-abled individuals such as quadriplegics and people with visual impairments. Although there is a growing body of research showcasing the use of breath activity as an interaction paradigm} 
 \changedelete{the field is still under-explored, particularly in terms of \emph{evaluating user experience} and \emph{designing breathing gestures}.}
\change{As an interaction mechanism, it is subtle (nearly invisible, suitable for discreet input where overt movements or voice commands are inappropriate), inherently hands- and eyes-free, and universally accessible, offering potential for individuals with severe motor impairments like quadriplegia \cite{esiyok2020software}, potentially complementing established methods like Sip-and-Puff interfaces \cite{mougharbel2013comparative,da2019development,jones2008sip}. While prior research has explored breath interaction \cite{elsahar2021study, esiyok2020software, burr2023breathtures, sra2018breathvr}, often using air pressure sensors or focusing on rate monitoring, these approaches typically require sensor placement near the mouth or nose, which can reduce social acceptability (e.g., Sip-and-Puff systems). In contrast, the design and user experience of discrete breathing gestures detected via wearable bio-impedance sensing, without requiring sensors near the face, remains largely underexplored.}

\changedelete{We address this gap by exploring a \emph{breathing-based gesture vocabulary}, designed to mimic the mental model of a  ``hand click''. }
\change{We address this gap by investigating foundational breathing gestures analogous to button clicks, sensed using upper-body bio-impedance.} We focus on "breathing clicks" (single, double, triple) for three reasons. First, they offer a familiar interaction metaphor. Second, clicks are fundamental, rapid activation primitives in conventional interfaces (mouse clicks: 150-250 ms \cite{keyboardkings}). While hands-free alternatives like eye-dwelling (400-800 ms \cite{majaranta2014eye}), voice commands (500-1000 ms \cite{murad2018design}), or head movements (500-1000 ms \cite{minakata2019pointing}) exist, exploring the feasibility and usability of click-style gestures implemented through breathing patterns sensed via bio-impedance is a novel area of exploration. Third, these simple clicks can serve as modular building blocks for potentially more complex breathing-based interactions.


Our work here focuses on the research question: \emph{how can we design a system for detecting user-friendly and comfortable breathing gestures during static and dynamic activities?} We specifically have \emph{three} contributions:

\begin{enumerate}
\item \emph{Gesture Detection System Development}. The technical implementation of the ``iBreath’’ wearable system for sensing breathing gestures using bio-impedance. The system effectively distinguishes breathing gestures from regular breathing by monitoring upper body bio-impedance variations caused by lung air volume changes (over 90\% accuracy).

\item \emph{Gesture Development and Evaluation}.  
We designed and evaluated novel breathing gestures (single, double, triple clicks) across static (sitting, lying) and dynamic (walking) scenarios through two user studies (total n=34). The first (n=8) assessed detection robustness, while the second (n=26) examined user experience, learnability, and accuracy. Results show the gestures were easy to learn and perform, with high detection accuracy (user-dependent models >90\% with ~140s training).  \change{Users preferred single-clicks and disliked triple-clicks. }

\item \emph{Design guidelines}. 
We synthesize our findings into eight concrete guidelines for designing bio-impedance-based breathing gesture interaction systems,  \change{covering hardware, algorithms, and interaction design. }
\end{enumerate} 


\changedelete{Our proposed system is significantly cheaper and supports higher mobility compared to the existing breathing detection systems, supports sensing gestures rather than breathing rates, and is easily learnable by the users. While our findings offer promising implications for designers seeking new input and interaction modalities for computer interfaces, the requirement for skin-attached electrodes is a key research challenge worth investigating for broader adoption. The proposed interaction is not meant as a replacement for mouse clicks in intensive tasks but rather \textbf{a complementary modality for multitasking or hands-free scenarios,} such as gaming, surgery, or cooking.} 
\change{Our work advances the state-of-the-art by demonstrating the feasibility of detecting discrete breathing gestures (not just respiratory rate) using low-cost, mobile bio-impedance sensing and evaluating their usability. We show these gestures can be learned quickly and recognized accurately. However, the practical adoption hinges on overcoming challenges like the comfort and integration of skin-attached electrodes, a key area for future work. \textbf{We position iBreath not as a universal replacement for mouse clicks or touch, but as a complementary modality enabling interaction in specific multitasking or hands-free scenarios where conventional input is impractical or unavailable (e.g., sterile environments like surgery, assistive technology for users with motor impairments, discreet control of wearables, or hands-occupied tasks like gaming or cooking)}}.
Our system could be packaged as a plug-and-play solution, simplifying implementation and enabling designers to explore breath gesture design. Beyond gesture detection, it could also be adapted to identify breathing patterns for health monitoring, sports training, and other applications.

\section{Related Work}
\label{sec: related_work}
\changedelete{Research on human-computer interaction mechanisms has been extensively explored in recent years, encompassing both the development of innovative interaction interfaces and the design of compact hardware configurations to enhance user interactions. }

\change{This section summarizes relevant literature on hands-free interaction methods, breathing-based interaction paradigms, and associated sensing technologies, highlighting existing gaps and opportunities addressed by our work.}

\subsection{Hands-Free Interactions}

\todofu{shorten+focus pros/cons of current methods}
Hands-free interaction is an advanced interaction strategy apart from traditional tactile button press to realize input through different body parts or touch-less gesture detection, enabling hands to execute other tasks.
Existing works like \cite{murad2018design} and \cite{monteiro2021hands} have already stated the importance of hands-free interaction fueled by the aspects of using digital devices for multi-task activities secondary to operating by hand.
Especially areas with high hygiene requirements like medical and health care \cite{monteiro2021hands}.
Over the past two decades, the human-computer interaction and wearable computing communities have extensively researched the promising potential of hands-free interaction solutions. These innovative modalities leverage various body parts such as eyes \cite{GazeTap, majaranta2014eye, valtakari2021eye, tonsen2017invisibleeye,liu2022non}, mouth \cite{ashbrook2016bitey, jingu2023lipio, ClenchClick2022}, ears \cite{roddiger2021earrumble}, nose \cite{polacek2013nosetapping, plotkin2010sniffing}, feet \cite{zaman2018interactive, minakata2019pointing} or shoulders \cite{geissler2024head} as the primary means of interaction.
For example, eye tracking serves as an effective method for interaction in the VR application scenario \cite{keskin2023potential}; 
The wearable eye-tracking device typically necessitates immobilizing the head to enhance data accuracy, with the primary drawback being the constraining aspect of such setups \cite{valtakari2021eye}. The computer-vision-based solution is the most used hardware setup, requiring more computation and hardware resources in data processing compared to sensor-based solutions, bringing challenges to resource-constraint wearable devices.
A discreet hands- and eyes-free input using the tensor tympani was demonstrated in EarRumble \cite{roddiger2021earrumble}, while only 43.2\% of the respondents had the ability to control it, which prevents it from being widely used. 
The LipIO \cite{jingu2023lipio} proposed a solution enabling the lips as the interaction interface by attaching a flexible PCB with electrodes to lips, this hardware setup suffers from low social acceptance.
A similar limitation also appears in the work ClenchClick \cite{ClenchClick2022}, which explored teeth-clenching-based target selection using the EMG sensing modality requiring attaching the electrodes on the face.
In addition, the foot-based interaction mechanisms were demonstrated in existing works \cite{zaman2018interactive,minakata2019pointing,tsai2024gait}, which provided a hands-free interaction solution for sterile environments like operating rooms, while this solution could limit the mobility of the users as the foot is occupied during the interaction.
Furthermore, these methods could not be suitable for people with disabilities, like quadriplegics.
\change{Each modality presents trade-offs regarding robustness, user comfort, social acceptance, and applicability to different user groups and contexts \cite{bian2022state,bian2024body}. Our work explores breathing as an alternative that is potentially more subtle and universally accessible than many of these options.}

\subsection{Breathing as an Interaction Paradigm}

The breath activity is controlled by the autonomic nervous system and the central nervous system, meaning that respiration can be consciously controlled by the subject, working as an explicit control signal in the HCI area \cite{cowley2016psychophysiology,zhang2019breath,lukic2022breathing,kusabuka2020ibuki,kim2020controlling}.
\change{Among the established breathing-based interaction techniques, the Sip-and-Puff system is notably prominent in assistive technologies \cite{da2019development,jones2008sip}. These systems translate distinct sip and puff actions, detected via air pressure sensors usually integrated into a mouthpiece, into control signals for devices like wheelchairs or consumer electronics. While highly valuable for specific user groups, Sip-and-Puff typically relies on mouth-based apparatus resulting in low social acceptance and focuses on binary or simple sequential actions rather than nuanced breathing patterns.}
Breathing interactions have also been explored in gaming, providing additional biofeedback to enhance immersive experiences \cite{sra2018breathvr,tatzgern2022airres,arroyo2009exploring,kim2020controlling,chen2025breath}. For instance, BreathVR \cite{sra2018breathvr} employed breathing as a physiological input channel, enabling natural and engaging gameplay interactions. Nonetheless, deliberate use of specific breathing techniques or patterns as direct control mechanisms remains limited \cite{elsahar2021study,esiyok2020software,burr2023breathtures,han2020development}. Elsahar et al. \cite{elsahar2021study} investigated recognizing four distinct breathing patterns using air pressure sensors for augmentative and alternative communication, achieving approximately 91.97\% accuracy; however, the user experience aspects were not evaluated. Han et al. \cite{han2020development} demonstrated the feasibility of using respiration-modulated signals from photoplethysmography sensors to toggle switches but evaluated only a single, simple gesture.
Similarly, BREATHTURES \cite{burr2023breathtures} proposed five breathing gestures—holding breath, deep abdominal breathing, deep thoracic breathing, abdominal staccato breathing, and thoracic sighing—as distinct input modalities. However, these gestures were potentially confusable with normal breathing patterns, such as relaxed breathing, thereby affecting detection accuracy. Furthermore, their user experience evaluation was limited to a very small sample of seven participants, and the reliability of their method was not thoroughly reported.
Although breathing offers considerable promise as an interaction modality, comprehensive research exploring practical gesture designs, robustness in detection, and thorough user experience evaluations is lacking. Our work addresses these critical gaps, advancing breathing-based interaction techniques significantly.

\subsection{Breath-Gesture Detection Technology}
Many existing works proposed promising solutions for breath activity monitoring including contact and remote methods \cite{ali2021contact} with different sensing modalities. \cref{table:related_work} lists examples for the related work,like acoustical \cite{gong2022breathmentor,hundia2019breathin, hou2022buma,shih2019breeze}, temperature \cite{basra2017temperature,wei2009novel}, pressure \cite{yu2019single,jin2014home}, IMU \cite{angelucci2023imu,liaqat2019wearbreathing}, RFID \cite{wang2022exhibit,wang2021retype}, mmWave radar \cite{choi2009remote} and bio-impedance \cite{lee2015integrated,sel2020wrist}.
However, most of them were designed primarily for monitoring the respiration rate.
In the design of HCI applications, it can be useful to pay attention to more speciﬁc details of respiration than respiration rate alone \cite{cowley2016psychophysiology}.
Besides, the social acceptance and robustness should be considered in the design of the HCI applications. 
In existing breath gesture detection works, the most commonly utilized technologies are acoustical, air pressure, and temperature sensors \cite{elsahar2021study, esiyok2020software,sra2018breathvr}. However, these methods present several drawbacks in HCI applications. For instance, they typically need to be positioned close to the mouth or nose, which is not socially favored in everyday environments, and the recording of acoustical signals may invade privacy and be affected by ambient noise. Additionally, IMU-based approaches are often influenced by the user's movements. These challenges might be more effectively managed with an RFID-based approach \cite{wang2022exhibit}, although it necessitates the installation of an antenna.\changedelete{Alternatively, bio-impedance sensing measures lung conductance, which correlates with the internal air volume, and only requires two electrodes on the body. This method is robust and discreet, making it ideal for wearable technology. Moreover, bio-impedance sensing can monitor breathing in sports for example. 
Although breath rate detection is widely explored}
Bio-impedance sensing offers a unique window into internal physiological and muscular activity, enabling more accurate and robust recognition of subtle or low-motion human activities that are difficult to capture with other sensors, which is already implemented in many human activity recognition research \cite{liu2024imove,liu2024ieat,liu2024iface,liu2025contrastive,liu2023ieat}.
\change{Our work selects bio-impedance sensing due to its direct correlation with lung volume changes} \cite{grenvik1972impedance}, \change{which is crucial for distinguishing detailed breath gestures. While IMUs offer skin-contact-free sensing and PPG provides high convenience through existing wearables } \cite{ liaqat2019wearbreathing,fusco2015extract}, \change{bio-impedance provides a more direct physiological measure of respiratory effort. Specifically, we employ an under-armpit (mid-axillary line) electrode placement, informed by literature suggesting this configuration can offer robust signals across postures and potentially minimize arm movement interference compared to traditional chest or more motion-prone wrist placements} \cite{ sel2020wrist, lee2015integrated, wang2014robust}. \change{While this choice aims to balance signal quality for gesture detection with discreet wearability, we acknowledge and further discuss (Section \ref{sec: discussion}) the ongoing challenges of long-term electrode comfort and mitigating self-touching artifacts. Thus, while bio-impedance for rate tracking is established, its application to detailed breath gesture detection for HCI, particularly with under-armpit sensing and a focus on usability and design guidelines, remains an under-explored area, which iBreath addresses.}

\begin{table}[h!]
\centering
\footnotesize
\caption{Examples for existing Breathing Detection Technology}
\label{table:related_work}
\begin{tabular}{ l l l l}
\hline
\textbf{Year-work} & \textbf{Application} & \textbf{Sensor} & \textbf{Location} \\ \hline
2009-\cite{choi2009remote} & breath rate & mmWave radar & 0.5 meters from chest \\ \hline
2018-\cite{sra2018breathvr} & breath gesture & Zephyr BioHarness & Chest \\ \hline
2019-\cite{hou2022buma} & breath rate & Microphone & Nose \\ \hline
2019-\cite{shih2019breeze} & breath rate & Microphone & Around belly \\ \hline
2019-\cite{liaqat2019wearbreathing} & breath rate & IMU & Wrist \\ \hline
2020-\cite{sel2020wrist} & breath rate & Bio-impedance & Wrist \\ \hline
2021-\cite{wang2022exhibit} & breath gesture & RFID tag & Chest area \\ \hline
2022-\cite{wang2021retype} & breath gesture & RFID tag & Chest area \\ \hline
2022-\cite{gong2022breathmentor} & breath rate & Microphone& In front of the user \\ \hline
2022-\cite{hundia2019breathin} & breath rate & Microphones & Near the sleeping person \\ \hline
2023-\cite{angelucci2023imu} & breath rate & IMU & Thorax, Abdomen and Lower back \\ \hline
2024-iBreath & breath gesture & Bio-impedance & Under armpits \\ \hline
\end{tabular}
\end{table}
 
\subsection{Research Gap and Novelty Statement}

Previous research focused on sensing breathing rates using radars, microphones, bio-impedance, and IMU sensors \cite{choi2009remote, hou2022buma, shih2019breeze, liaqat2019wearbreathing, sel2020wrist, angelucci2023imu}. However, gesture detection differs significantly from breathing rate sensing, as it requires monitoring both \emph{the speed and depth} of breaths rather than just inhalation and exhalation frequency. Despite the growing interest in breathing as a hands-free interaction paradigm \cite{elsahar2021study,esiyok2020software,burr2023breathtures,han2020development}, literature on \emph{breathing gesture detection} remains scarce. Therefore, researchers and designers would benefit from clear guidelines for designing practical and robust breathing gestures prioritizing user-friendliness. Prior work exploring breathing gesture detection \cite{sra2018breathvr, wang2022exhibit, wang2021retype} typically used commercial biosensors (e.g. the Zephyr BioHarness) and RFID tags to minimize signal noise. Our custom-built prototype offers a significant cost advantage~\footnote{e.g. Costs: our prototype = 40 USD, the Zephyr BioHarness = 800 USD} and provides open access to the data required for gesture detection which was blocked in commercial APIs we tried. Bio-impedance sensing is also better for mobility compared to RFID tags~\footnote{Participants must be in the same room with the antenna when using RFID tags. However, our design could possibly use a mobile phone to receive and process the data via bluetooth.}. To our knowledge, there is no previous research on simulating clicking gestures using bio-impedance.

Our novelty lies in two key areas: 1) extending the use of bio-impedance through a \emph{novel algorithm} to sense breathing gestures rather than only rates, and 2) designing gestures, particularly clicks, from repetitive blocks while extensively investigating user experience to provide concrete design guidelines. \changedelete{The designed gestures could be used as vocabulary to design more complex gestures}\change{The designed gestures could be used as modular building blocks for potentially more complex breathing-based interactions.} Closing this gap is crucial as well-designed breath gestures can benefit nearly everyone, especially those with disabilities.

\begin{figure}[t]
\includegraphics[width=1.0\linewidth ]{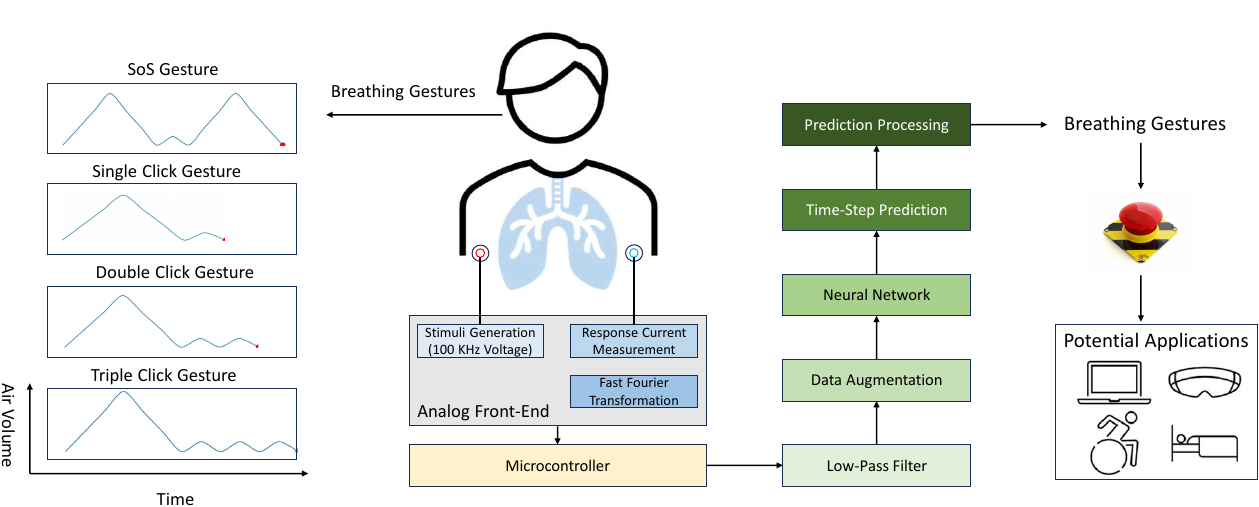}
\caption{System Design of iBreath (The iBreath system includes the breathing gesture design, wearable hardware for bio-impedance measurement, and the breathing gesture recognition algorithm using neural networks )}
\label{fig: system_design}
\end{figure}

\begin{table}[h!]
\centering
\caption{Summary of designed gestures using iBreath. \cref{fig: system_design} shows the graphs.}
\begin{tabular}{c|c}
\hline
Gesture Name & Description \\
\hline
\textbf{SOS (customized help gesture}) & One fast breath between two deep ones\\
\textbf{Single-click (\cone)} & One deep breath followed by one quick breath.\\
\textbf{Double-click (\ctwo)} & One deep breath then two quick breaths\\
\textbf{Triple-click (\cthree)} & One deep breath then three quick breaths \\
\hline
\end{tabular}
\label{tbl:gestures}
\end{table}


\section{Part 1: Breathing Gesture Detection}

\change{"iBreath" detects breathing gestures from the regular breathing signal using bio-impedance. This section presents the system design and the validation study.} \cref{fig: system_design} provides an overview of the iBreath system design composed of three parts: 1) basic breath gestures, 2) wearable hardware to monitor bio-impedance\changedelete{changes} \change{signal}, and 3) the breathing gesture recognition algorithm using neural networks. The core principle is that lung air volume changes upper body bio-impedance, with inhalation increasing and exhalation decreasing it. These fluctuations are linked to variations in rhythm, speed, and intensity. We use those variations to detect unusual breathing patterns as gestures. 


\subsection{Component 1: Basic Breath Gestures}
Unlike hand gestures, there are no standard breathing gestures for interaction in current research. Thus, we used two familiar variations in breathing patterns: \emph{breath depth} and \emph{breath speed} to design our gestures. We specifically used deep and slow breaths vs shallow and fast ones as our building blocks. Our goal was to design gestures that are easy and memorable for the users while recognizable by the detection algorithm from regular breathing. 

Existing research shows deep breathing improves mood and reduces stress \cite{perciavalle2017role}. Therefore, we began our designed breathing gesture with deep breathing to alleviate user fatigue and avoid confusing the gesture with regular breathing. We incorporated fast breathing next, as its rapid and sharp nature differs from regular breathing and is time-efficient, reducing latency during model inference. The algorithm can recognize the gestures as the deep breath results in higher impedance and a fast breath is shorter in duration than a regular breath. Table \ref{tbl:gestures} summarizes the description of the four gestures we designed. We mainly designed clicks to have building blocks for more complex gestures and because it is widely spread. We added the SOS signal to validate that the system can support custom gestures distinct from clicks.


\subsection{Component 2: Wearable Hardware} \label{sec:hardware}

As iBreath is a wearable device, size and power consumption are crucial considerations in hardware development. iBreath monitors intentional breathing patterns with bio-impedance sensors and two electrodes under the armpits, detecting lung air volume changes. \change{Notably, electrode placement plays a big role in stability. A study evaluating multiple placements found that a configuration along the mid-axillary lines (essentially under the armpits) produced the most robust respiration signals across different postures} \cite{wang2014robust}.
\cref{fig: system_design} illustrates the architecture of the iBreath hardware setup, which consists of two primary components: the Analog Front-End (AFE) and the microcontroller. The  AFE  is connected to the microcontroller nRF52840 via the SPI interface, which controls the measurement procedure. Additionally, measurement results can be transmitted to a terminal through a Bluetooth interface. In this study, we configured the stimuli as a 100 kHz alternating voltage with a peak-to-peak amplitude of 50 mV. The bio-impedance sampling rate is set at 20 Hz. The user's body interfaces with the AFE via wet Ag/AgCl electrodes. The operating current of iBreath is approximately 20 mA. A compact 500 mAh lithium battery powers the system, sufficient for about one day of operation. The cost of the hardware setup is around 40 US dollars. Further details about the technical implementation of the sensing module is shown in \cref{sec: circuit}.


\subsection{Component 3: Breathing Gesture Detection Algorithm}\label{sec:comp3algo}
The breath gesture detection algorithm comprises \emph{three} key components (details in \cref{appendix:algorithm}): 1) data augmentation: to simulate real-world noise, 2) a neural network: \changedelete{to assign several predictions per gesture}\change{to predict the gesture from the sensor data}, and 3) post-processing of prediction results: \changedelete{to assign single class label per gesture}\change{to further improve the prediction accuracy based on prior knowledge}. The implementation parameters are summarized in \cref{app:params}.

\ressec{Data Augmentation (Increase Data \changedelete{Points}\change{Samples})} 
The data augmentation module improves the neural network's robustness against disturbances in bio-impedance measurements. These disturbances, termed system bias, include individual differences due to body fat and electrode placement variations. Additionally, bio-impedance changes vary with users' lung capacity and breathing habits (e.g. breathe from mouth or nose), and electrode movement introduces Gaussian noise. These factors complicate training a user-independent model, especially with limited subjects. To address this, we use data augmentation techniques to increase training samples by adding \textbf{three} types of noise to simulate expected disturbances, enhancing model generalizability: 1) \emph{shift}: simulates system bias to individual differences and location of the electrodes, 2) \emph{scale up/down}: simulates bio-impedance variation from lung capacity and breathing habits, and 3) \emph{gaussian noise}: simulating random noise from electrode movement due to human activity. The noise equations are reported in \cref{app:params}. Incorporating the three noise types has increased the number of training samples by \emph{fivefold}, significantly improving the network's robustness in supporting users' diversity in body fat and breathing habits.

\ressec{Neural Network (Time-Level Predictions)}

The ``BreathNet'' model we developed incorporates a sequence of one-dimensional convolutional (CNN1D), self-attention, long short-term memory (LSTM), and linear layers (see parameters' summary \cref{app:nerualparams}). The model training uses a cross-entropy loss function and the Adam optimizer, configured with a learning rate of 5e-4 and a batch size of 512. Since the predominance of the null class in the training samples, leads to class imbalance, class weights were adjusted to more heavily compensate for less frequent classes, thus addressing the disparity in training data sample distribution across classes.
Training was conducted over 100 epochs, incorporating early stopping with a patience of 30 epochs to prevent overfitting. The model's input comprises two channels: the magnitude and phase of the bio-impedance. To accommodate the varying durations of the four breathing gestures and individual differences in gesture execution, a large window size of 100 was configured to capture all relevant gesture features. \change{A small slide step size of 5 was employed to maintain a high prediction update rate,}  ensuring that the prediction frequency closely matches the time-step-based prediction approach. The output from BreathNet is a time-step-level prediction result (i.e. several possible predictions per gesture). A summary of the network's structure and parameters is provided in \cref{fig:model_summary}. 

To enhance the performance of iBreath, an \textbf{optimization algorithm of time-step level prediction} was developed (see summary in \cref{fig: data_flow,sec: alorthm}). The algorithm is suitable for \textbf{real-time}, time-step-level prediction optimization. This algorithm incorporates three sequential strategies ~\footnote{Full description of the standard strategies provided in \cref{app:params}} based on the following prior knowledge of breathing gestures:
\begin{enumerate}
    \item Each breathing gesture should yield multiple time-step level predictions, ensuring that these non-null time-step level predictions are consecutive and exceed a single time-step level prediction in length. $\rightarrow$ Algorithm uses a \emph{low-pass strategy}: reduce noise and eliminate outliers. 
    \item If the window samples are incomplete, double, triple, and SoS clicking gestures may be identified as a single click gesture. Similarly, a triple-click gesture could be recognized as a double-click gesture. $\rightarrow$  Algorithm uses a \emph{front-follows-back strategy}: correct false positives from sliding window.
    \item There should be only one kind of breathing gesture performed in continual non-null predictions. $\rightarrow$  Algorithm uses a \emph{majority-rule strategy}: detect consistent predictions over a defined time range. Most time-step latency comes from here but it does not affect event-level latency.
\end{enumerate} 

\ressec{Post-Processing of Prediction Results (Event-Level Predictions)}

In real-time prediction, we define a buffer of size ten to store up to ten previous predictions for applying these strategies. The buffer length is estimated based on the sliding window's step size, the sensor's sampling rate, and the duration of breathing gestures. In the experimental setup for this work, the buffer stores predictions from the past five seconds. This duration accommodates nearly all breathing gestures typically performed by most individuals.
Normally, the final decision from the optimization algorithm is made immediately when a null-class prediction takes place indicating one breathing gesture is finished.
Hence, this does not incur additional latency at the event-level prediction except the time consumption of additional computing.
The performance of the time-step level prediction optimization algorithm is discussed in \cref{sec:s1_result}. In real-time interaction scenarios, the breathing gesture command should only be issued after a complete gesture has been performed. Therefore, an event-level prediction is derived from the time-step level predictions. Specifically, a gesture event is identified when the last time-step prediction was non-null and the current time-step prediction transitions to null class. The result of the event-level prediction is determined by a majority vote from these previous consecutive time-step level predictions. Between two non-null event-level predictions, the gesture event is predicted to be null. 

\begin{figure}[t]
\includegraphics[width=1.0\linewidth ]{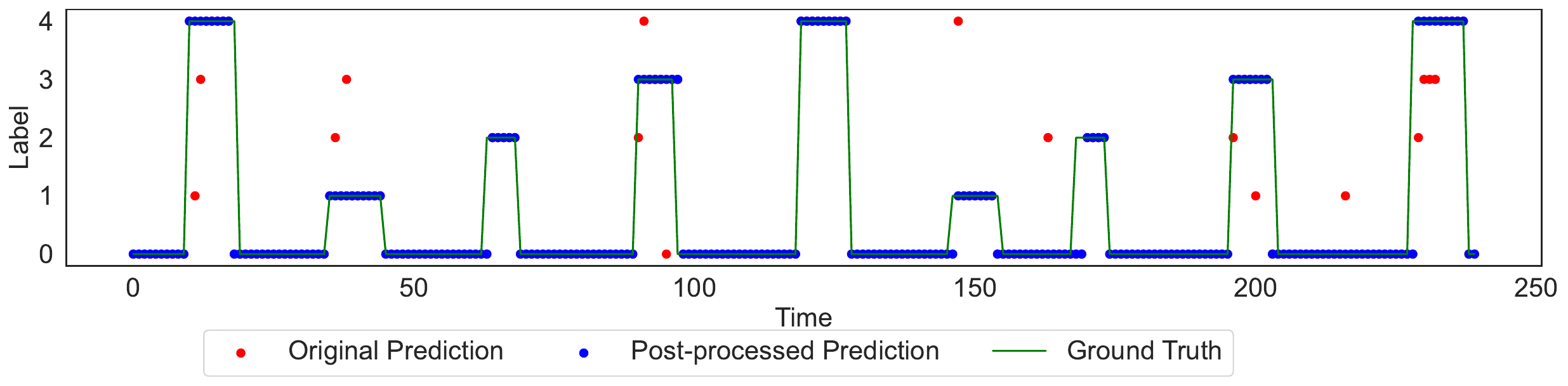}
\caption{Time-step level prediction optimized by the post-processing method (there are multiple different original predictions (Red points) from the neural network when the user only performs one breathing gesture, the outlier original prediction can be removed through post-processing method, as the results shown in red points)}
\label{fig: prediction_optimized}
\end{figure}

\begin{figure}[t]
\includegraphics[width=1.0\linewidth ]{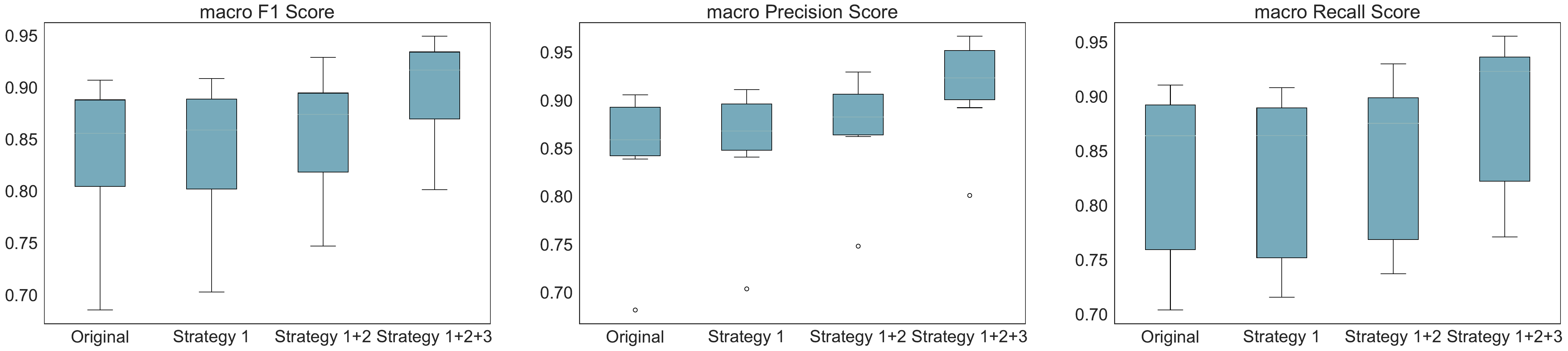}
\caption{The performance of optimization algorithm on time-step level prediction results (Strategy 1: Low-Pass Strategy, Strategy 2: Front follows back strategy, Strategy 3: Majority rule strategy. These three strategies are used successively and additively to the original result)}
\label{fig: box_opitimization_result}
\end{figure}

\subsection{Validation Study: \change{System Robustness and Feasibility}} \label{sec:validation_study_1} 

\change{To evaluate the robustness and feasibility of the iBreath system, we conducted a validation study. This study assessed the system's performance in detecting four distinct breathing gestures across three common free-living scenarios. We utilized a cross-validation approach to analyze sensor data processed by our BreathNet algorithm, focusing on both time-step and event-level prediction accuracy.}

\subsubsection{Methodology} \label{sec:study1_methodology}
\change{We conducted a validation study involving eight participants (mean age 29.21 years, Min=26, Max=32, SD=2.32). These individuals, recruited via snowball sampling, were students or researchers without known respiratory diseases and new to breathing gestures. The 40-minute sessions were conducted in a typical meeting room, allowing for normal activities. Participation was voluntary and uncompensated.}

\change{The study employed a within-subject experimental design. Participants wore the iBreath hardware (detailed in Section \ref{sec:hardware}, placement shown in \cref{fig: sensing_principle}) and interacted with a custom JavaScript Web App. This app, using Google Chrome's BLE API, provided a visual training animation (a rising/falling red ball) for gesture synchronization, real-time visualization of bio-impedance data streamed via Bluetooth, and logged labeled breathing signals for offline analysis. This setup permitted participants to move freely.}

\change{The experiment manipulated two independent variables: 1) \emph{scenario} representing the user's activity (3 conditions: walking, sitting, and lying) and 2) \emph{gesture} representing the breathing gesture (4 conditions: SOS, single click, double click, and triple click).} \change{The procedure began with participants providing informed consent and being fitted with the iBreath device. After the experimenter demonstrated the training animation and task, participants completed experimental blocks for each scenario (fixed order: sitting, lying, walking). Within each ~10-minute scenario block, the Web App prompted participants to perform each of the four gestures 15 times in a randomized order (totaling 60 trials per scenario, 180 trials per participant). During sitting and lying scenarios, participants could move naturally but were asked to maintain their general posture. A two-minute rest was provided between scenario blocks. Sensor data collected from these trials were subsequently processed using our BreathNet algorithm, and system performance was assessed using cross-validation methods focusing on both time-step and event-level predictions.}

The performance of the iBreath system was assessed using various metrics at both the time-step and event levels. The metrics were: F1 score, recall, precision, and accuracy. We employed two cross-validation methods: leave-one-person-out (LOPO) and leave-one-scenario-out (LOSO), to ensure comprehensive evaluation across different conditions. Table \ref{tab:s1_result_summary} summarizes the metrics and methods. To address class imbalance, results are presented using both macro and weighted average methods across various classes. 


\subsubsection{Results}
\label{sec:s1_result}

\cref{tab:s1_result_summary} provides a summary of the first study. 
Whether on the leave-one-person-out (LOPO) or leave-one-scenario-out (LOSO) test, the iBreath system achieved an average recognition accuracy of over 95\% at the time-step level and over 93\% at the event-level prediction, showing effectiveness in practical applications.

\ressec{Raw Signal Measurement Result}

\begin{figure*}[!h]
\centering
  \begin{subfigure}[b]{0.49\textwidth}
  \includegraphics[width=1.0\linewidth]{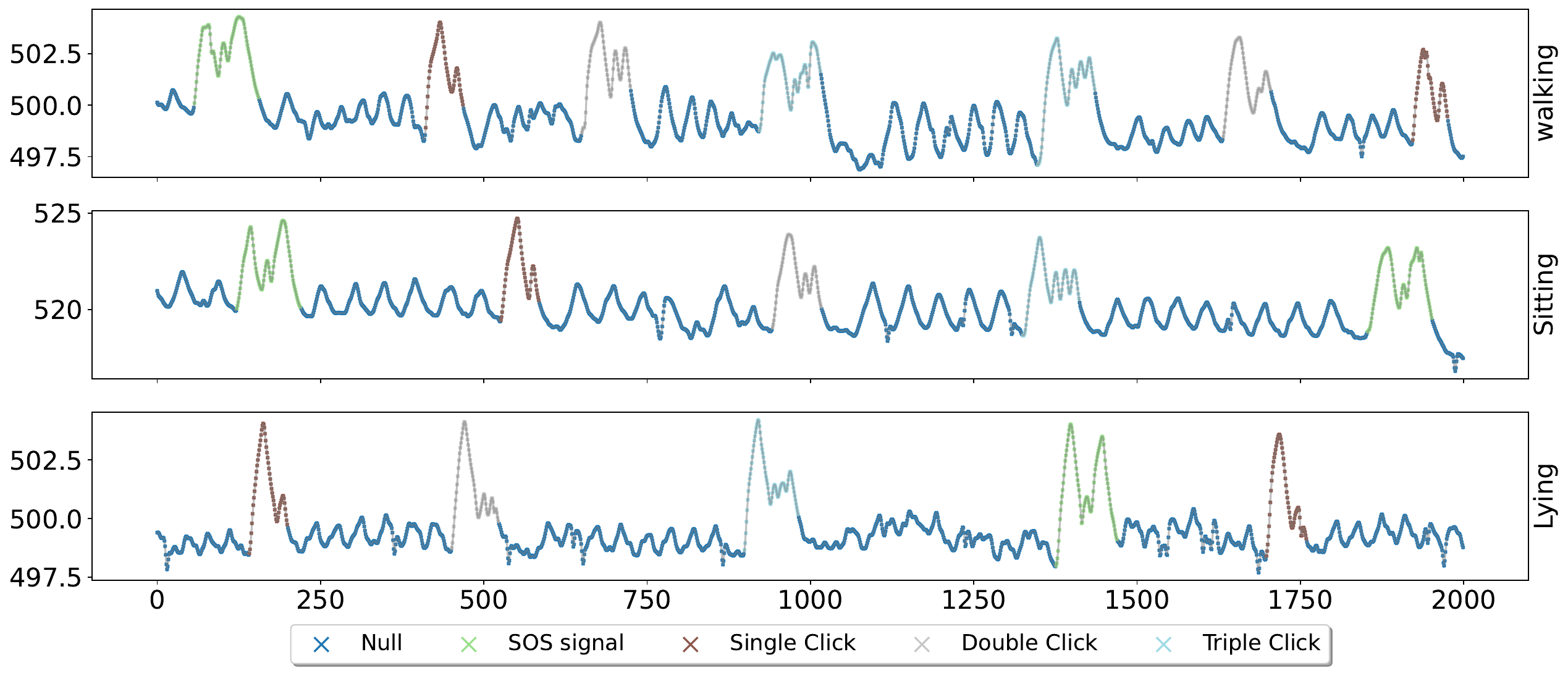} 
  \caption{Raw magnitude signal from subject 4}
  \label{fig:raw_signal_magnitude_4}
  \end{subfigure}
  \begin{subfigure}[b]{0.49\textwidth}
  \includegraphics[width=1.0\linewidth]{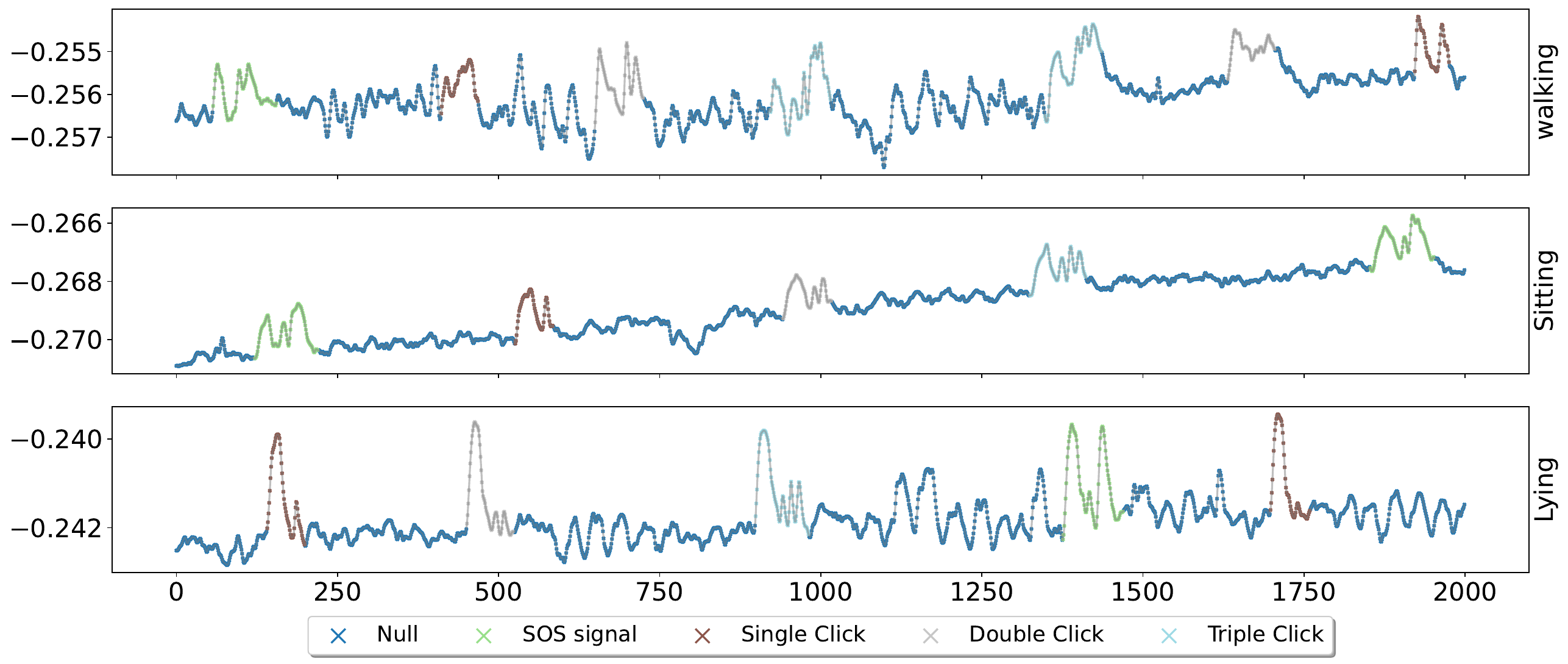} 
  \caption{Raw phase signal from subject 4}
  \label{fig:raw_signal_phase_4}
  \end{subfigure}

\begin{subfigure}[b]{0.49\textwidth}
  \includegraphics[width=1.0\linewidth]{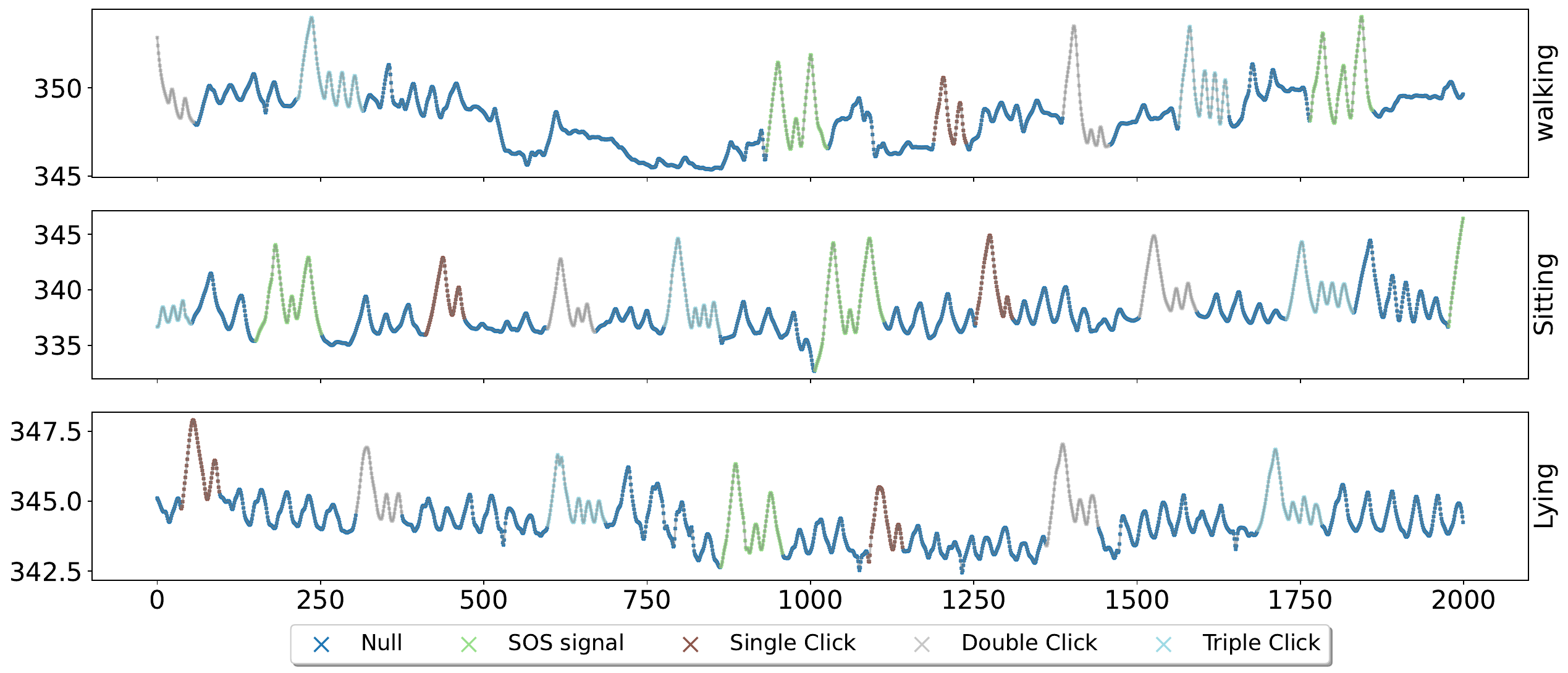} 
  \caption{Raw magnitude signal from subject 7}
  \label{fig:raw_signal_magnitude_7}
  \end{subfigure}
  \begin{subfigure}[b]{0.49\textwidth}
  \includegraphics[width=1.0\linewidth]{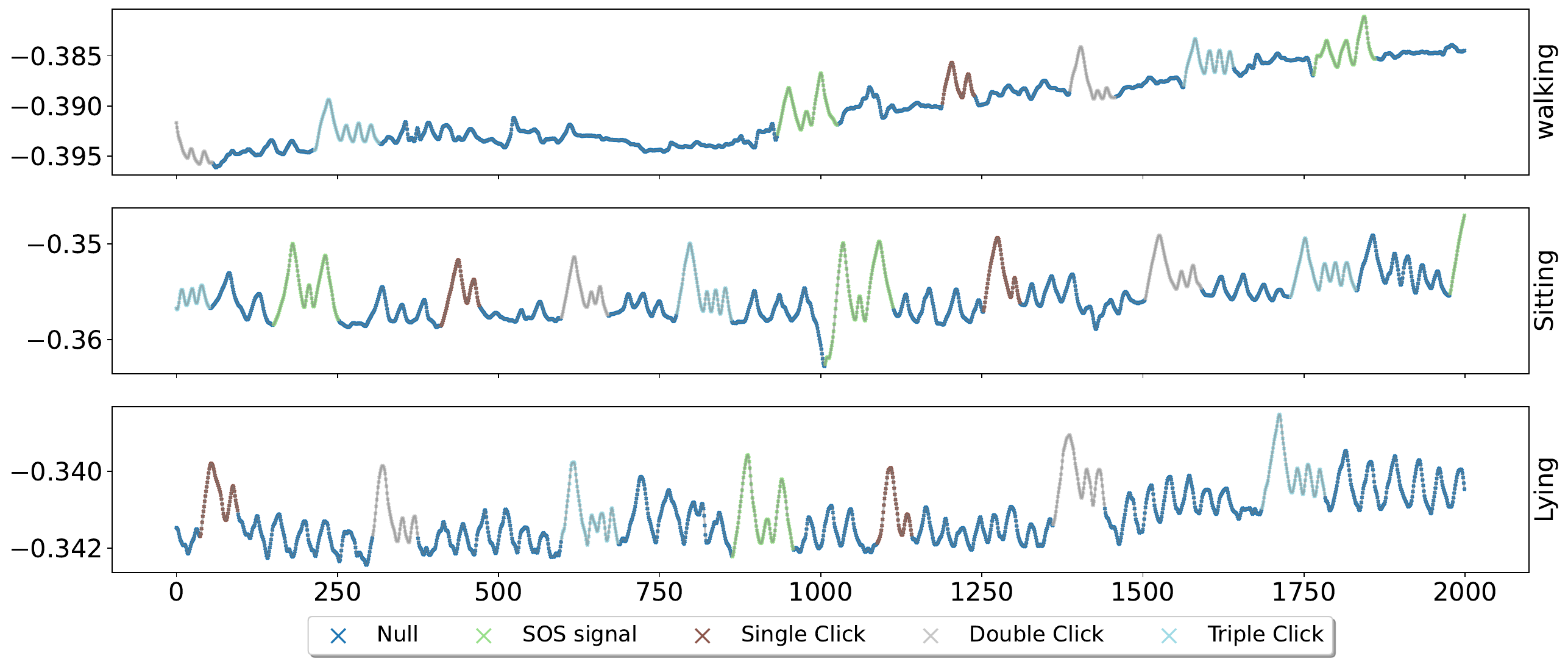} 
  \caption{Raw phase signal from subject 7}
  \label{fig:raw_signal_phase_7}
  \end{subfigure}
\caption{Raw bio-impedance signals from breathing activity}
\label{fig: raw_signal}
\end{figure*}

\cref{fig: raw_signal} displays bio-impedance signals caused by breathing activities in three different scenarios from two subjects. It is evident that the iBreath system is capable of measuring both regular and intentional breathing gestures effectively in sitting and lying positions. However, the quality of the signals monitoring regular breathing shows noticeable degradation in the walking scenario, particularly for Subject 7, as illustrated in \cref{fig:raw_signal_magnitude_7}. Despite this, breathing gesture signals are still discernible. Across all scenarios, the variations in bio-impedance signals induced by breathing gestures are greater than those caused by regular breathing. Notably, the magnitude channel provides a clearer indication of breathing gestures compared to the phase channel. Furthermore, for Subject 4, the magnitude of the signal following fast breaths after deep breathing activities is higher than that during regular breathing, as shown in \cref{fig:raw_signal_magnitude_4}. Although Subject 7 does not show a marked difference in magnitude between fast and regular breathing, the variations can still be differentiated by the frequency of breathing. These observations confirm that iBreath can effectively capture distinctive features of breathing gestures from regular breathing activities.

\begin{table}[!t]
\centering
\footnotesize
\caption{Result Summary of Study 1}
\label{tab:s1_result_summary}
\begin{tabular}{c c c c c c c }
\hline

\hline
Prediction Level&Cross-Validation Method & Average Method & F1 Score (\%) & Precision (\%)& Recall (\%)& Accuracy (\%)\\
\hline

\hline
\multirow{4}{*}{Time Step Level} & \multirow{2}{*}{Leave One Person Out} & Macro & 89.65 ± 5.24 & 91.50 ± 4.97 & 88.62 ± 6.80 & \multirow{2}{*}{95.28 ± 2.59 }\\
                                 && Weighted & 95.21 ± 2.64 & 95.38 ± 2.58 & 95.28 ± 2.59 & \\
                                &\multirow{2}{*}{Leave One Scenario Out} & Macro & 90.17±2.36 & 91.94 ± 0.39 & 88.75 ± 4.36 & \multirow{2}{*}{95.57 ± 0.96}\\
                                && Weighted & 95.50 ± 1.04 & 95.54 ± 1.04 & 95.57 ± 0.96 & \\

\hline          

\hline

\multirow{4}{*}{Event Level} & \multirow{2}{*}{Leave One Person Out} & Macro & 92.54 ± 7.48 & 91.67 ± 7.55 & 94.11 ± 7.07 & \multirow{2}{*}{93.60 ± 6.43 }\\
                                 && Weighted & 93.74 ± 6.21 & 94.41 ± 5.60 & 93.60 ± 6.43 &  \\
                                &\multirow{2}{*}{Leave One Scenario Out} & Macro & 93.34 ± 0.15 & 92.41 ± 0.55 & 94.73 ± 0.15& \multirow{2}{*}{93.78 ± 0.14}\\
                                && Weighted & 93.82 ± 0.18 & 94.21 ± 0.25 & 93.78 ± 0.14 &  \\

\hline          

\hline

\end{tabular}

\end{table}

\ressec{Prediction Optimization Result}
The optimization method for post-processing the time-step level predictions from BreathNet significantly enhances accuracy, as illustrated in \cref{fig: prediction_optimized}. Initially, the BreathNet produced some false positive predictions that appeared as outliers, which were effectively removed using a low-pass filtering strategy. Additionally, in cases like the triple-click gesture, the algorithm initially misidentified it as a single-click and double-click before correctly identifying it as a triple-click. This error was rectified using the Front-Follows-Back strategy, which updates false positives in light of subsequent correct predictions.
Moreover, while direct time-step level predictions were inconsistent—even when only one type of breathing gesture was performed—the majority rule strategy corrected false predictions that frequently appeared as isolated anomalies within sequences of consecutive non-null predictions. Despite several time-step discrepancies between the non-null predictions and the ground truth, these did not adversely affect the event-level predictions.
The overall performance of the optimization algorithm on time-step level predictions is further detailed in \cref{fig: box_opitimization_result}. Here, we observe significant improvements in macro F1-score, precision, and recall. Notably, Strategy 3 proved to be the most effective in enhancing these metrics.

\begin{figure*}[!h]
\centering
  \begin{subfigure}[b]{0.61\textwidth}
  \includegraphics[width=1.0\linewidth]{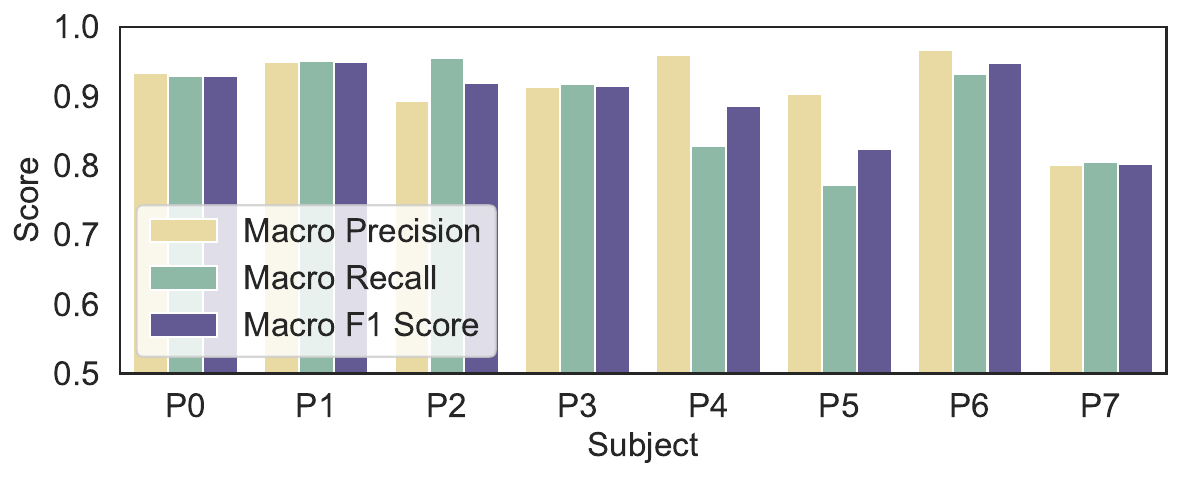} 
  \caption{Performance on each subject (TSL)}
  \label{fig:s1_lpo_bar_ts}
  \end{subfigure}
  \begin{subfigure}[b]{0.37\textwidth}
  \includegraphics[width=1.0\linewidth]{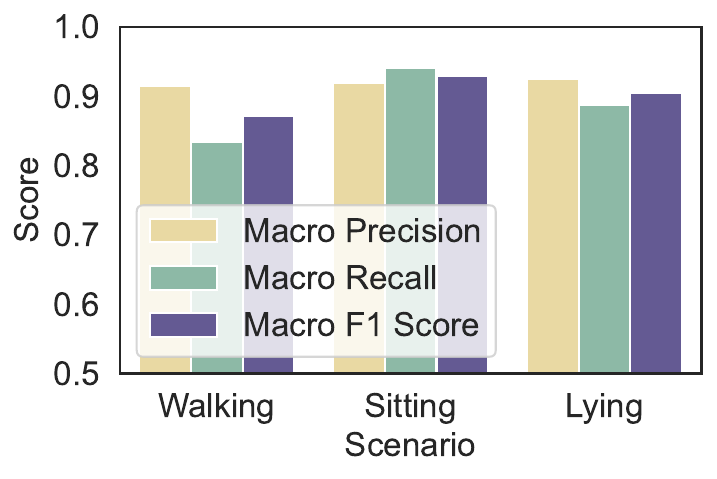} 
  \caption{Performance on each scenario (TSL)}
  \label{fig:s1_lso_bar_ts}
  \end{subfigure}

\begin{subfigure}[b]{0.61\textwidth}
  \includegraphics[width=1.0\linewidth]{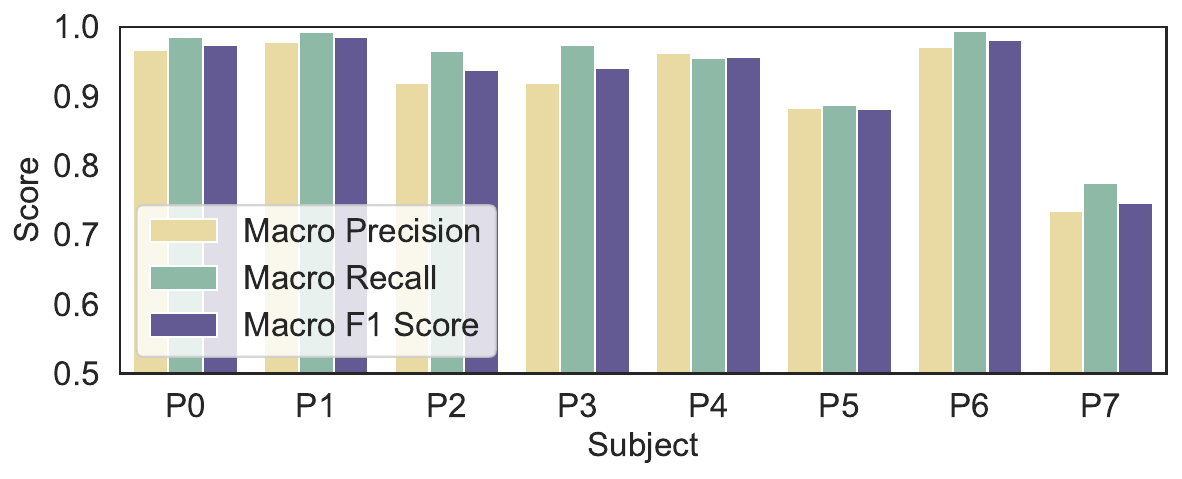} 
  \caption{Performance on each subject (EL)}
  \label{fig:s1_lpo_bar_el}
  \end{subfigure}
  \begin{subfigure}[b]{0.37\textwidth}
  \includegraphics[width=1.0\linewidth]{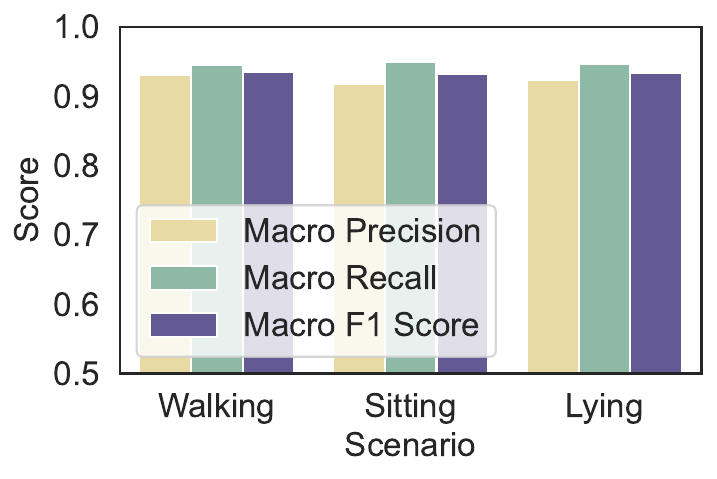} 
  \caption{Performance on each session (EL)}
  \label{fig:s1_lso_bar_el}
  \end{subfigure}
\caption{Detailed prediction result of each subject and each scenario with the different cross-validation method and different level prediction. (TSL: Time-Step Level, EL: Event Level)}
\label{fig:s1_detailed_result_bar_ts}
\end{figure*}

\begin{figure*}[!h]
\centering
  \begin{subfigure}[b]{0.24\textwidth}
  \includegraphics[width=1.0\linewidth]{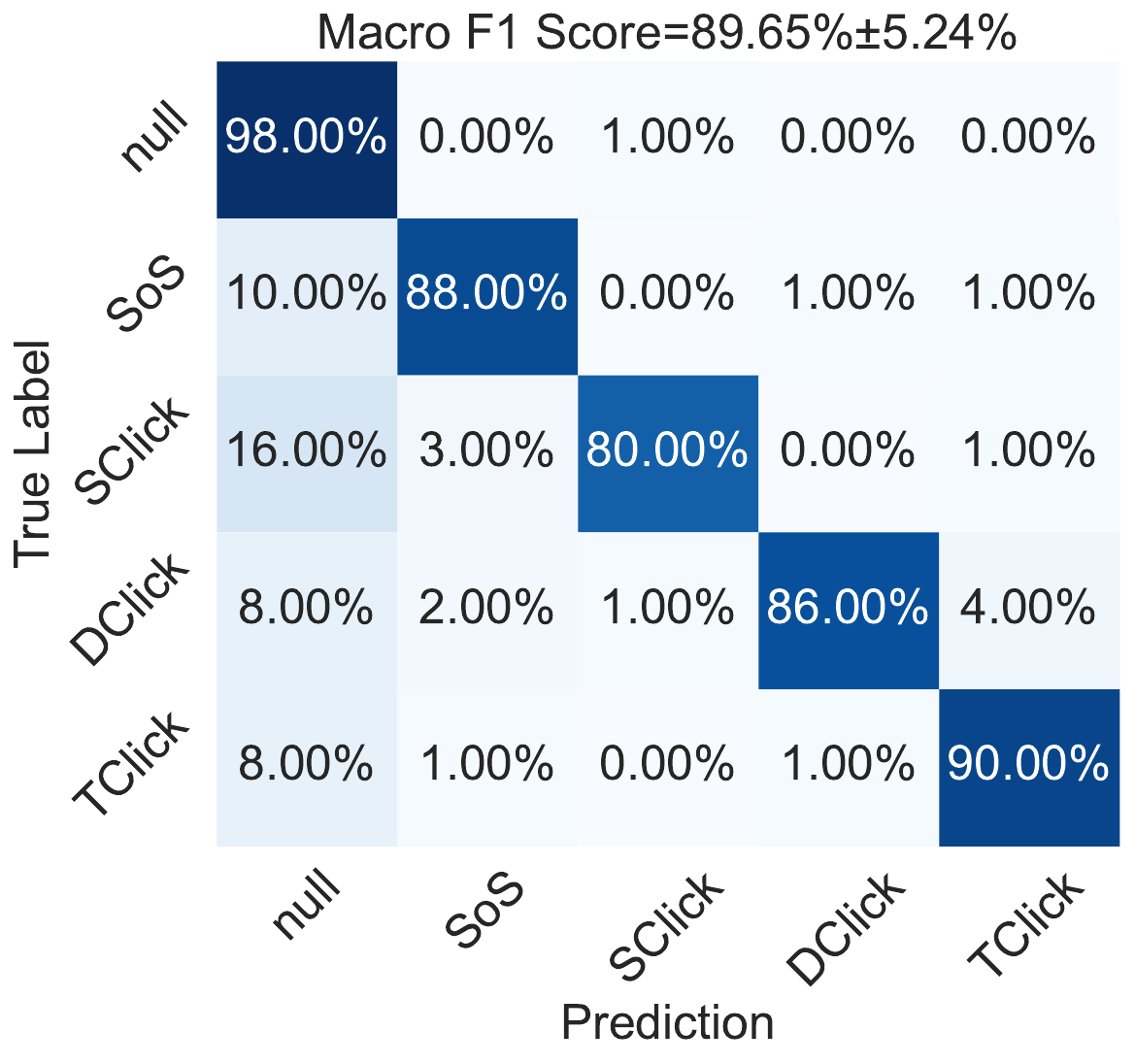} 
  \caption{TSL-LOPO}
  \label{fig:s1_cm_lpo_tsl}
  \end{subfigure}
  \begin{subfigure}[b]{0.24\textwidth}
  \includegraphics[width=1.0\linewidth]{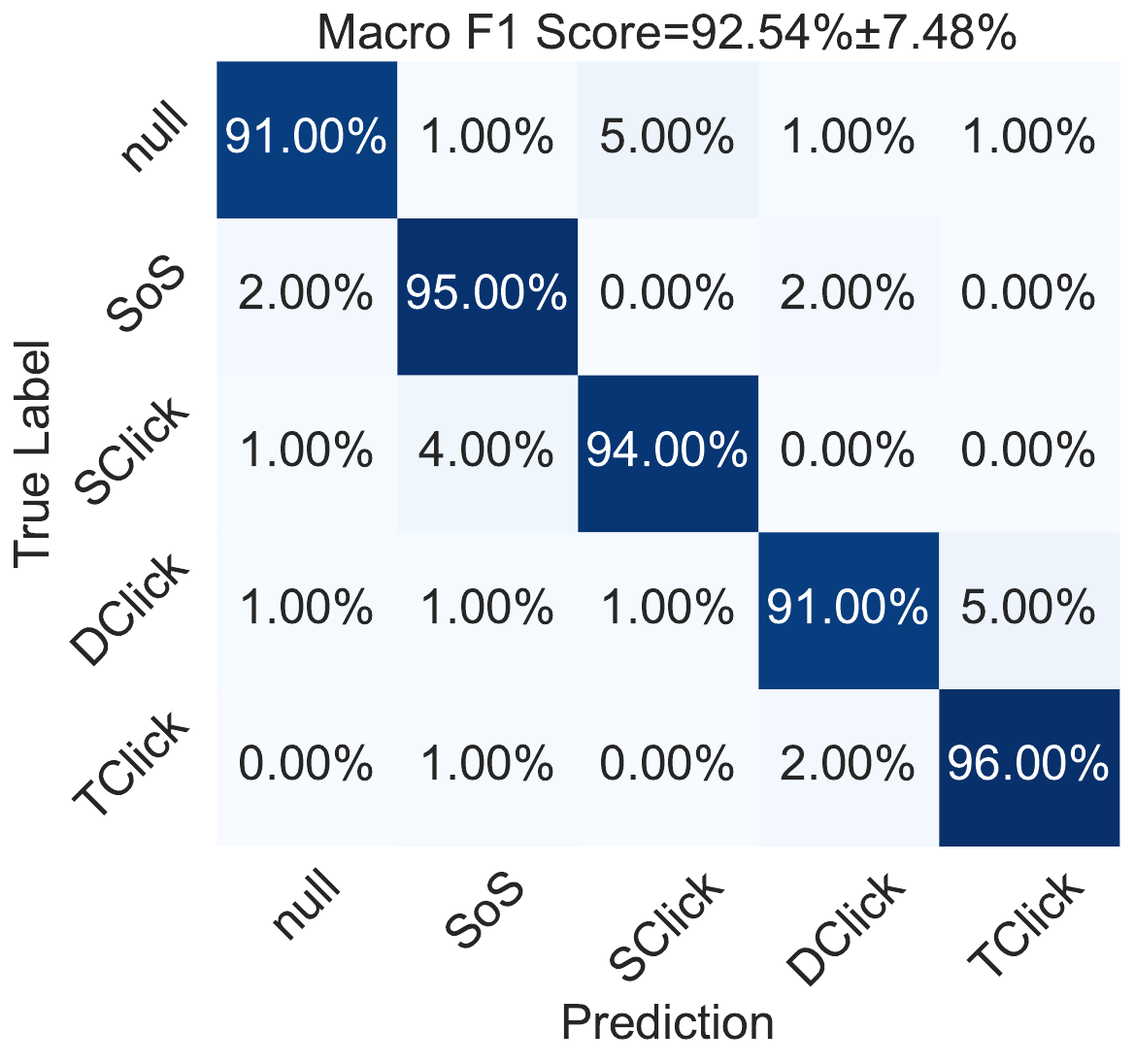} 
  \caption{EL-LOPO}
  \label{fig:s1_cm_lpo_el}
  \end{subfigure}
\begin{subfigure}[b]{0.24\textwidth}
  \includegraphics[width=1.0\linewidth]{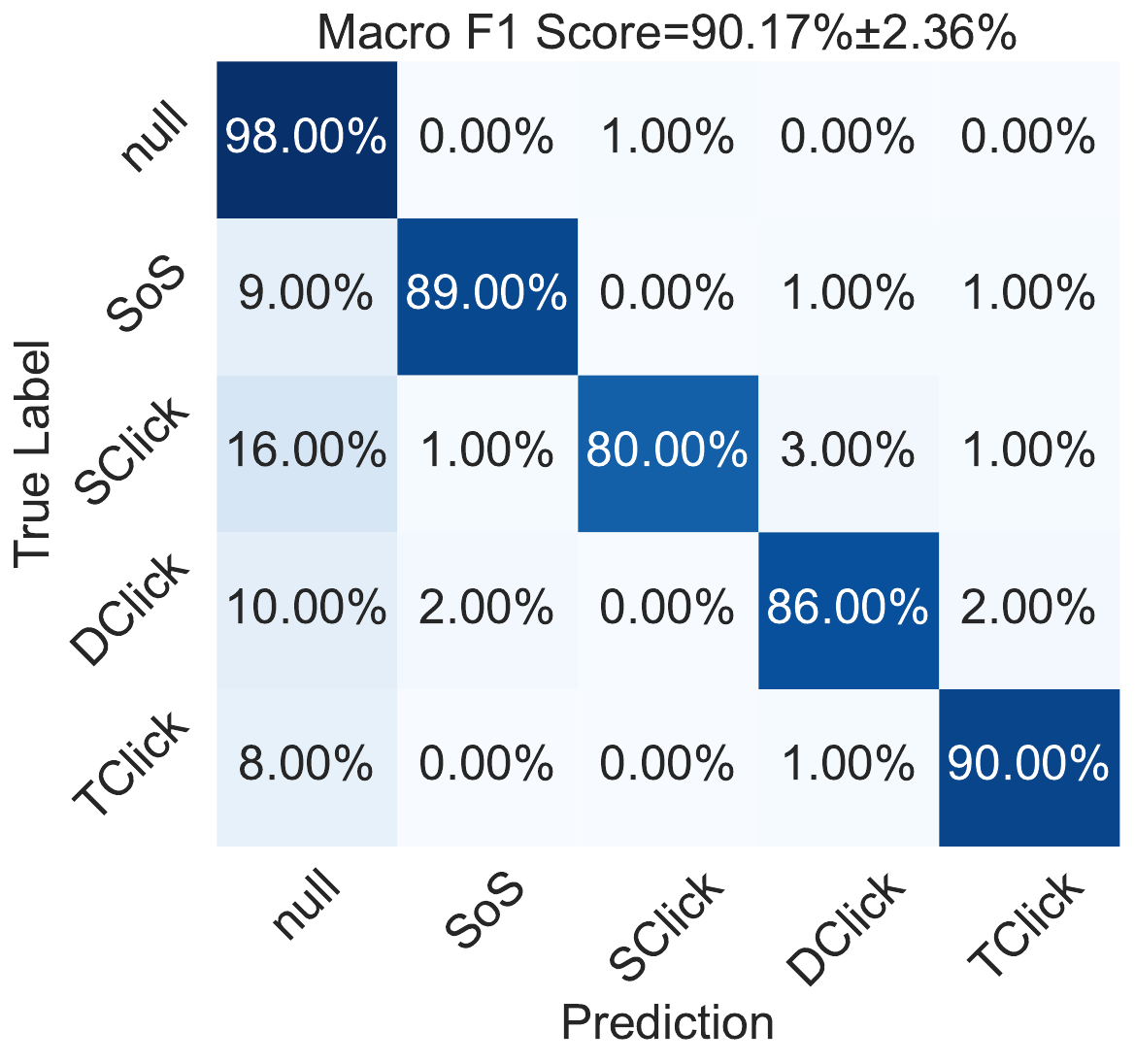} 
  \caption{TSL-LOSO}
  \label{fig:s1_cm_lso_tsl}
  \end{subfigure}
\begin{subfigure}[b]{0.24\textwidth}
  \includegraphics[width=1.0\linewidth]{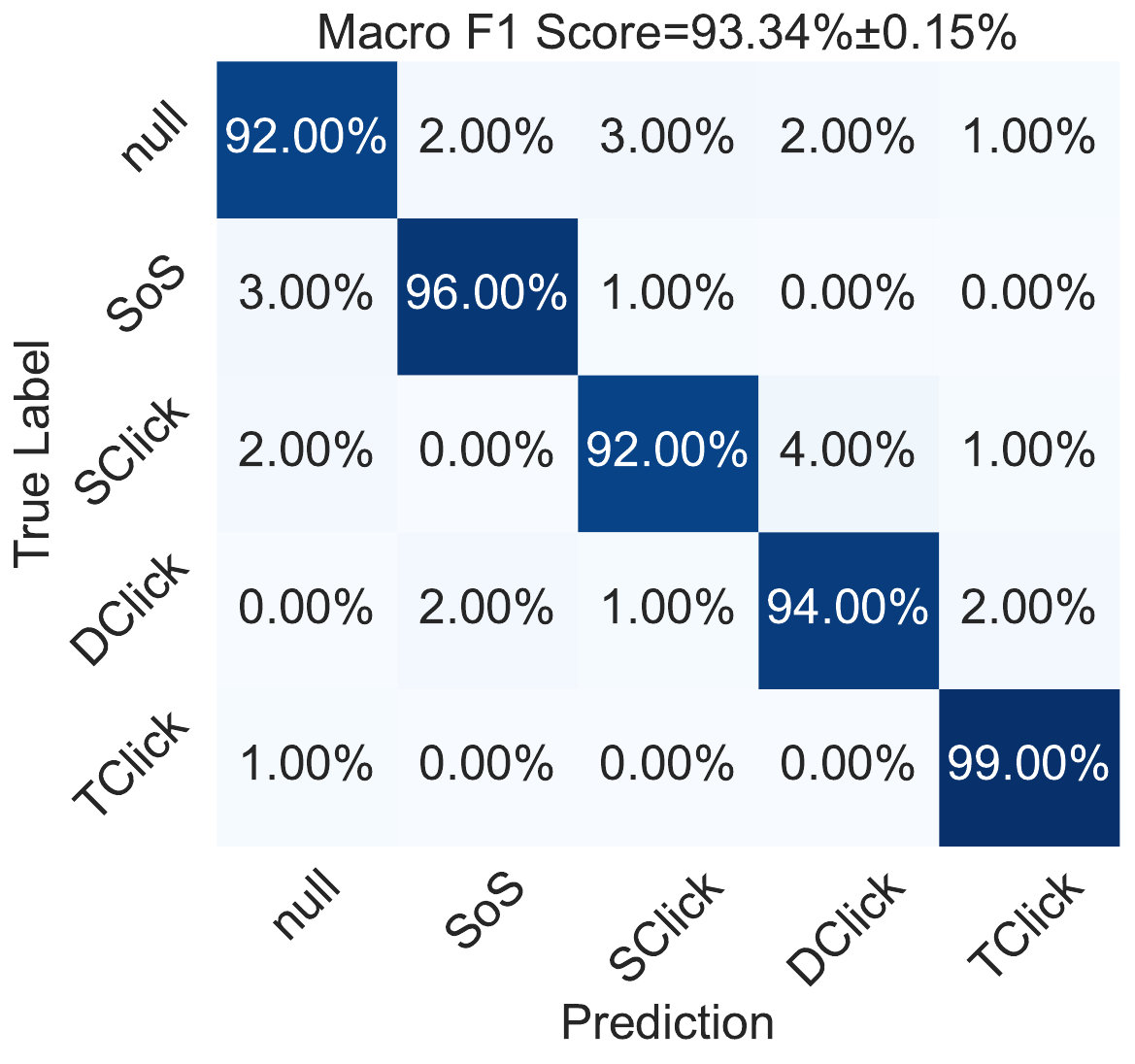} 
  \caption{EL-LOSO}
  \label{fig:s1_cm_lso_el}
  \end{subfigure}
\caption{Joint Confusion Matrix of iBreath with the different cross-validation method and prediction level.(\textbf{LOPO}: Leave-One-Person-Out, \textbf{LOSO}: Leave-One-Scenario-Out, \textbf{TSL}: Time-Step Level prediction, \textbf{EL}: Event Level prediction, 
\textbf{SClick}: single click gesture, \textbf{DClick}: double click gesture, \textbf{TClick}: triple-click gesture)}
\label{fig:s1_cm}
\end{figure*}

\ressec{Accuracy based on Prediction Level and Cross-Validation}
\change{We evaluated iBreath's accuracy at both time-step and event levels using LOPO and LOSO cross-validation, revealing robust performance overall}.

\change{At the Time-Step Level, optimized predictions (confusion matrices in \cref{fig:s1_cm_lpo_tsl} for LOPO and \cref{fig:s1_cm_lso_tsl} for LOSO) achieved strong macro F1 scores of 89.65\% (LOPO) and 90.17\% (LOSO). While BreathNet generally distinguished different breathing gestures effectively, the primary misclassifications occurred between the null class and breathing gestures, particularly single-clicks. Individual participant and scenario breakdowns (\cref{fig:s1_lpo_bar_ts,fig:s1_lso_bar_ts}) show that most subjects' gestures were recognized with macro precision, recall, and F1 scores consistently above 90\%. Lower scores for Subject 7 likely stemmed from feature similarity between their fast and irregular breathing patterns (\cref{fig:raw_signal_magnitude_7}), while the dynamic walking scenario unsurprisingly yielded the lowest scores in LOSO due to motion-induced bio-impedance noise. Performance in other, more static scenarios consistently exceeded 90\%.}

\change{The Event-Level Performance, critical for practical interaction as commands issue only upon full gesture completion, demonstrated even higher robustness. Event-level confusion matrices (\cref{fig:s1_cm_lpo_el} for LOPO, \cref{fig:s1_cm_lso_el} for LOSO) consistently show recall scores above 90\% for all gestures and the null class. Notably, this approach significantly reduced confusion between null and breathing gesture classes; initial time-step discrepancies (e.g., \cref{fig: prediction_optimized}) did not degrade these final event-level outcomes. The Triple-click gesture achieved the highest recognition rates, closely followed by the SOS gesture, across both LOPO and LOSO evaluations. Detailed data (\cref{fig:s1_lpo_bar_el,fig:s1_lso_bar_el}) confirm marked improvements over time-step predictions and consistently robust performance across all scenarios. This underscores iBreath's capability to effectively process breathing gestures for interaction in both dynamic and stationary conditions.}

\changedelete{\ressec{Impact of user activity on system performance}}

\changedelete{We calculated the precision and recall for each user activity (walking, sitting, and lying) while leaving one scenario out. The results demonstrated robust performance, with precision rates between 91.7\% and 93.1\%. Consequently, we did not further investigate potential statistical differences, as they were deemed minimal from an interaction perspective. We denote here precision with a (P) and recall with an (R). The values were as follows: walking activity (P= 93.1\%, R= 94.5\%), sitting activity (P= 91.7\%, R= 94.9\%), and lying activity (P= 92.3\%, R= 94.6\%).} 


\subsection{Takeaways}
Overall, the comprehensive results demonstrated the reliability and robustness of the iBreath based on bio-impedance sensing technology used for breathing gesture detection. 
We list below the key findings used to design the second user study in the next section.

\begin{enumerate}
    \item The iBreath system demonstrates high accuracy in detecting breathing gestures, achieving \change{recognition accuracy of} over 95\% \changedelete{recognition} at the time-step level and over 93\% at the event level, showcasing its reliability for practical applications.

    \item Both LOPO and LOSO cross-validation methods indicate robust performance across diverse subjects and scenarios, including dynamic activities, confirming the system's reliability.

    \item Event-level predictions excel with recall scores above 90\%, highlighting the system's efficacy in generating reliable interaction commands based on complete gesture recognition.

    \item The iBreath effectively distinguishes between various breathing gestures, with proposed strategic optimizations like Front-follows-Back significantly reducing misclassifications and improving time-step level prediction accuracy.

    \item The gesture recognition is highly robust (91.7 \% - 93.1 \%) across different user activities such as walking, sitting and lying.
    
\end{enumerate}

\section{Part 2: Evaluating the User Experience while using the Gestures}
\label{sec: second_study}

\insertfig{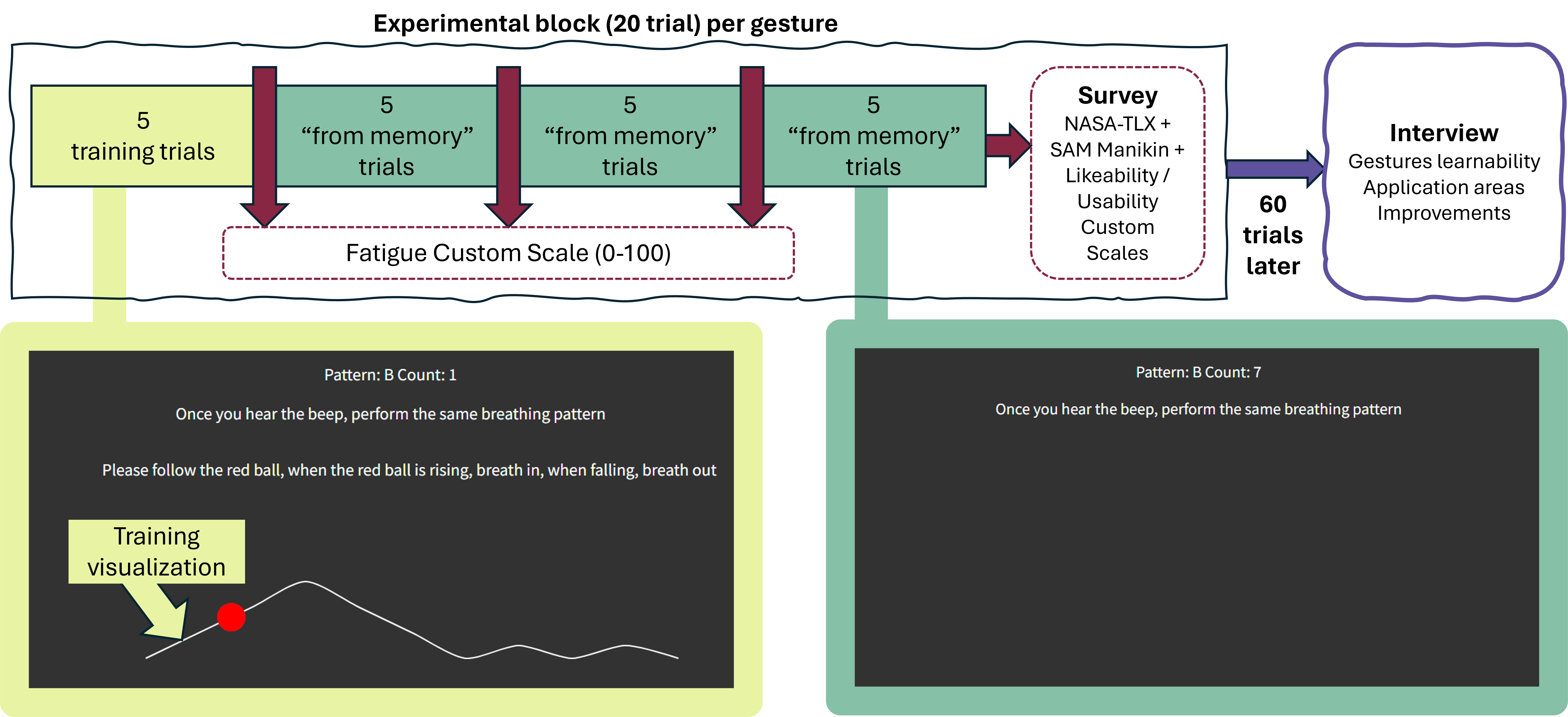}{fig:s2-method}{The experimental design for the study described in Part 2. The goal is to evaluate the usability and the detection accuracy of the three new breathing gestures: single click, double click and triple click. The screenshots show the apparatus. Figure \ref{fig: system_design} shows the training visualizations for the three gestures.}{1}{blocks showing the experiment order}

Based on the robustness demonstrated in our initial study, we proceeded to a larger user experience evaluation focusing specifically on the click-based gestures (single, double, triple), as they represent a fundamental and widely understood interaction primitive. The goal was to assess both user experience and detection accuracy of these newly designed gestures across a broader participant pool.

\subsection{Methodology}

\change{The experimental design for this user study is summarized in Figure \ref{fig:s2-method}. Ethical approval was obtained from the relevant institutional review board(s) prior to participant recruitment. We employed a within-subject design primarily investigating the impact of one independent variable: gesture type (3 conditions as defined in Table \ref{tbl:gestures}: single-click, double-click, and triple-click). For clarity, a "breath" in this context refers to a single sequence of inhalation followed by exhalation.}

\change{Our key dependent variables included user experience (assessed via questionnaires) and detection accuracy. Detection accuracy was further analyzed based on two conditions of another independent variable: the detection model type (user-dependent vs. user-independent algorithm). The specific metrics for detection accuracy were precision (the accuracy of positive predictions) and recall (the completeness of predictions).}

\change{The experiment involved a single lab session per participant. Each trial consisted of performing one breathing gesture (lasting approximately 10 seconds) cued by an auditory beep. An experimental block for each gesture type comprised 20 trials: the initial five for training using a visual animation (detailed in Section \ref{sec:study1_methodology}), followed by 15 trials where participants performed the gesture from memory. Participants were naive to the specific gestures before the training visualization. To mitigate order effects, the presentation sequence of the three gesture types was counterbalanced across participants using a balanced Latin square design. After completing all 20 trials for a given gesture type, participants filled out a questionnaire about their experience. Upon completion of all three gesture blocks, a short semi-structured interview was conducted.}

\subsubsection{Detection Model Training} In the user-dependent model, we used 14 trials from the current user to train the model. In the user-independent model, we used a total of 420 trials from 21 participants who provided 20 trials each. The data of the current user was not included. Both models were trained on the data of one gesture only and detected one class: gesture or not. We employed the same detection model, shown in \cref{fig:model_summary}, used in ``iBreath'' with a variation that we detect here one gesture class instead of four. 

\subsubsection{Survey Design } 

The survey’s goal is to evaluate the user’s experience while using the breathing gestures. The survey is found in \cref{sec: Questionnaire}. The survey is printed and comprises three tools: NASA-TLX \cite{hart2006nasa}, SAM-Manikin \cite{morris1995observations}, and two custom scales: likeability and usability of the gesture. We use NASA-TLX to understand how demanding the tasks is and the SAM-Manikin to capture the participant’s feelings while performing the gestures. The custom scales are direct questions to assess if the user likes the tested gesture and if they would use it as an interaction gesture. The NASA-TLX ranges are a continuous numerical scale [0,21] (0: very low and 21: very high).  
The SAM-Manikin values are an ordinal scale [1,9] (Happiness factor= 1: unhappy and 9: happy, Calmness factor= 1: calm and 9: excited, Control factor= 1: no control and 9: in control). The ranges of the custom likability and usability scales are also continuous numerical scales between [0,100]. The likability labels are (0: not at all, 100: love it) and the usability labels are (0: not at all, 100: yes, absolutely).
\subsubsection{Interview Design} 
We focused the interview on three aspects: 1) gestures’ complexity and learnability, 2) application areas for breathing gestures, and 3) areas for improvement. The full list of questions is provided in \cref{sec: Interview}. For the first category, we asked participants about their overall experience, the ease of understanding the training visualization and memorizing the gesture, and the required time for training. For the second category, we asked for application areas for breathing gestures, in general, to interact with environments, if participants would use the gestures to specifically click buttons, and their satisfaction with the gestures as an interaction technique. For the third category, we asked them about challenges and intuitive design aspects of the gestures, adoption challenges and any aspects they would like to change. 


\subsubsection{Apparatus} We use a similar web application to the one described in Section \ref{sec:study1_methodology}. During the experiment, the participants were asked to sit in front of a laptop wearing the iBreath hardware and follow the commands presented on the screen to complete the experiment. Users perform a gesture when the application beeps. There was a camera in front of the participants to record the whole study procedure and the interview.  


\subsubsection{Participants and Recruitment}
The experiment consisted of one lab session. We recruited 26 participants (10 females and 16 males) through snowball sampling and the institute’s mailing list. The mean age of the participants was 25.58 years ($min=21, max=34, std= 3.061$). The latest educational degree of the participants was as follows: 2 participants finished high school, 19 participants are bachelor holders, and 5 participants are master holders. Only 3 participants reported having prior breathing problems. However, none of the participants terminated their participation in the experiment. Three participants have tried breathing gestures before. We did not have filtration criteria for recruitment.  Participants were not compensated monetarily for their participation and they were explicitly instructed that they could terminate their participation at any point without consequences. On average,  the participant spent 35 minutes in the lab session.

\subsubsection{Procedure} Each lab session was moderated by one researcher. The researcher welcomed the participants and handed them the consent form. The consent form clearly stated potential risks from joining the experiment such as triggering patients with breathing problems or anxiety/panic disorders. We explicitly instructed them to terminate their participation if they experience any discomfort. The experiment started after signing the consent form. The experiment was video recorded to in case we needed to understand glitches with the null class during the post-processing. The participant filled out the demographics form. Afterwards, the researcher briefly explained the experimental procedure.
The researcher starts a random condition (gesture) for the participant and the apparatus controls the experimental flow (see Figure \ref{fig:s2-method}). 
After finishing the 20 trials for each gesture type, the software pauses and the researcher hands the participant a paper questionnaire comprised of NASA-TLX, SAM-Manikin, and the custom likability and usability scales. After filling out the questionnaire, the participant clicks a button to start the next gesture type repeating the same steps of 20 trials. After finishing the three gestures (a total of 60 trials), the researcher conducts a semi-structured interview with the participants to understand their interaction. The researcher thanks the participant afterwards and the session ends after storing the collected data in the designated GDPR-compliant institute repository. 


\subsubsection{Analysis} \label{sec:s2:analysis}
We used Bayesian Factor Analysis to compare the three gestures in terms of detection accuracy (precision and recall), user experience (using NASA-TLX, SAM Manikin), and explicit user preferences (likability and usability custom scales). We utilized Bayesian statistical tests instead of the standard frequentist inferential approach. These tests not only identify differences between groups but also assess the likelihood of similarity (null hypothesis). This translates to making claims that the gestures are similar in some aspects.  We analyzed the data using Jasp software \cite{jasp} where the corresponding Bayesian version of all appropriate regular frequentist tests was used \cite{rouder2009bayesian,morey2015bayesfactor,jasp}. 
\ressec{Utilized Statistical Tests} We employed a standard non-informative prior distribution with equally distributed probabilities across all conditions for all the Bayesian tests. All $B10$ values are the likelihood compared to the null hypothesis model \bfval{1}. We used a separate \emph{Bayesian Repeated Measures ANOVA} to compare the three gestures to each factor of the NASA-TLX, SAM-Manikin, likability and usability custom scales. The independent variable was the gesture type and the independent variable was the corresponding scale values. We conducted a total of 11 tests. We also used separate \emph{Bayesian Repeated Measures ANOVA} to check the gestures’ accuracy of prediction. However, we changed the input to test for two variations. The first variation was to test which detection model should be used for each gesture. Thus, we conducted 3 tests corresponding to each gesture with two independent variables: the model type (user-dependent and user-independent) vs. the model accuracy (precision and recall). The second variation was to compare the three gestures in terms of accuracy within each detection model separately. Thus, we conducted 2 tests corresponding to each detection model with two independent variables: the gesture type (single-click, double-click, and triple-click) vs the model accuracy (precision and recall). For all tests indicating significant differences between conditions, we followed by default T-tests with Cauchy Prior for post-hoc comparisons. 
\ressec{Data Processing} We only excluded participants from each factor separately when their data was incomplete. We report in the next section the number of participants excluded from each test and Table \ref{fig:s2-res-tbl-ux} summarize them. We used the raw values from all scales without further processing. The \emph{middle values} for the scales are: Nasa-TLX= 11, SAM-Manikin= 5, Likeability and Usability custom scales= 50. 

\ressec{Data Interpretation} For each factor, we analyze two aspects: 1) the overall trend by interpreting the mean value, and 2) the consistency of the trend across various conditions using the \bfnull value. In short, every Bayesian test has a \bfnull value. This value serves two purposes: 1) determining if there are differences between conditions, and 2) assessing the confidence level in this result (evidence strength). $BF_{10} > 1 $ indicates there is a difference between the conditions while $BF_{10} < 1 $ indicates the conditions are similar. The values of evidence strength are in descending order starting with the highest confidence: strong > moderate > weak. We want to mainly utilize results with strong and moderate strength. Weak ones correspond to anecdotal evidence so we cannot conclusively interpret it.  
\ressec{Qualitative Analysis} One researcher coded all questions into three closed codes: yes, no, and maybe. Each participant was coded into only one category. Afterwards, the researcher used open coding to analyse the challenges and the application areas. One participant could contribute to several codes. Afterwards, another researcher checked the codes and re-grouped them then reported the results.


\subsection{Results} \label{sec:s2results}


We present here the user experience evaluation of the three gestures: single click (\cone), double click (\ctwo), and triple click (\cthree). We focus on answering three research questions: 1) which gesture is more accurate to detect? 2) what was the user experience with the gestures? and 3) which gesture was favoured by the users? Here is a summary of the variable names as presented in the results section. \emph{Gesture type} (3 conditions:\cone, \ctwo, \cthree), \emph{Model} (2 conditions: user-dependent and user-independent),  \emph{Accuracy} (2 conditions: precision and recall). 


\begin{figure}
\insertsubfig{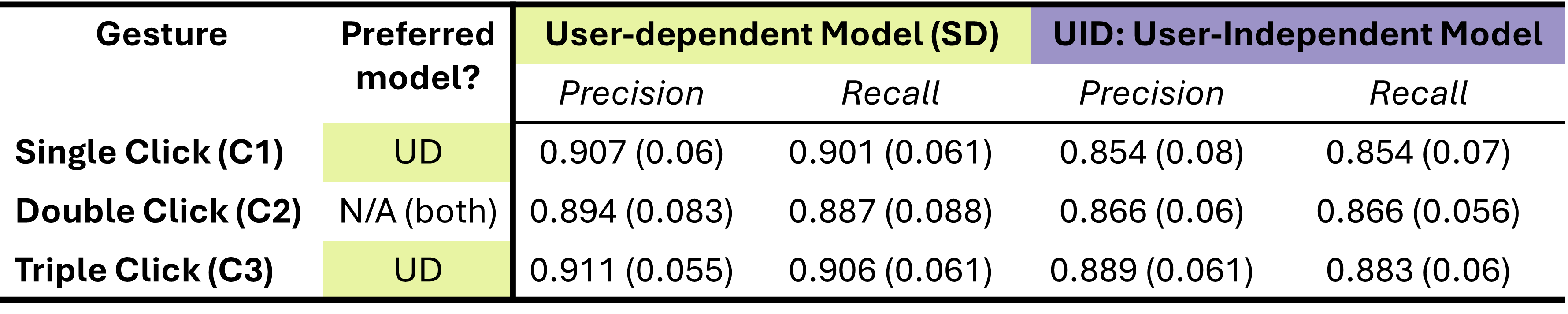}{fig:tbl-accuracy}{Summary of accuracy per model and gesture. Values are mean \% normalized.}{0.8}{graph}


\insertsubfig{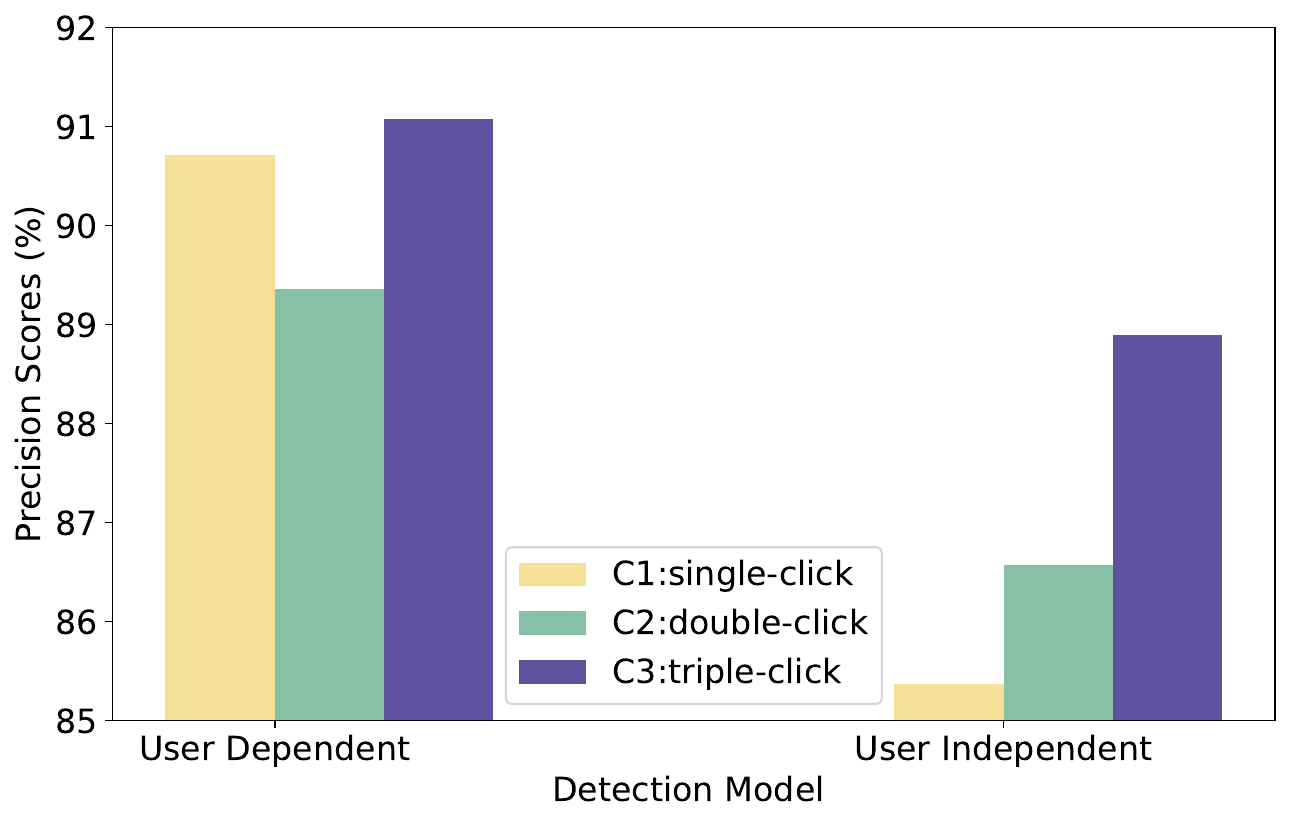}{fig:s2-res-precision}{Precision (\%).}{0.49}{graph}
\hfill
\insertsubfig{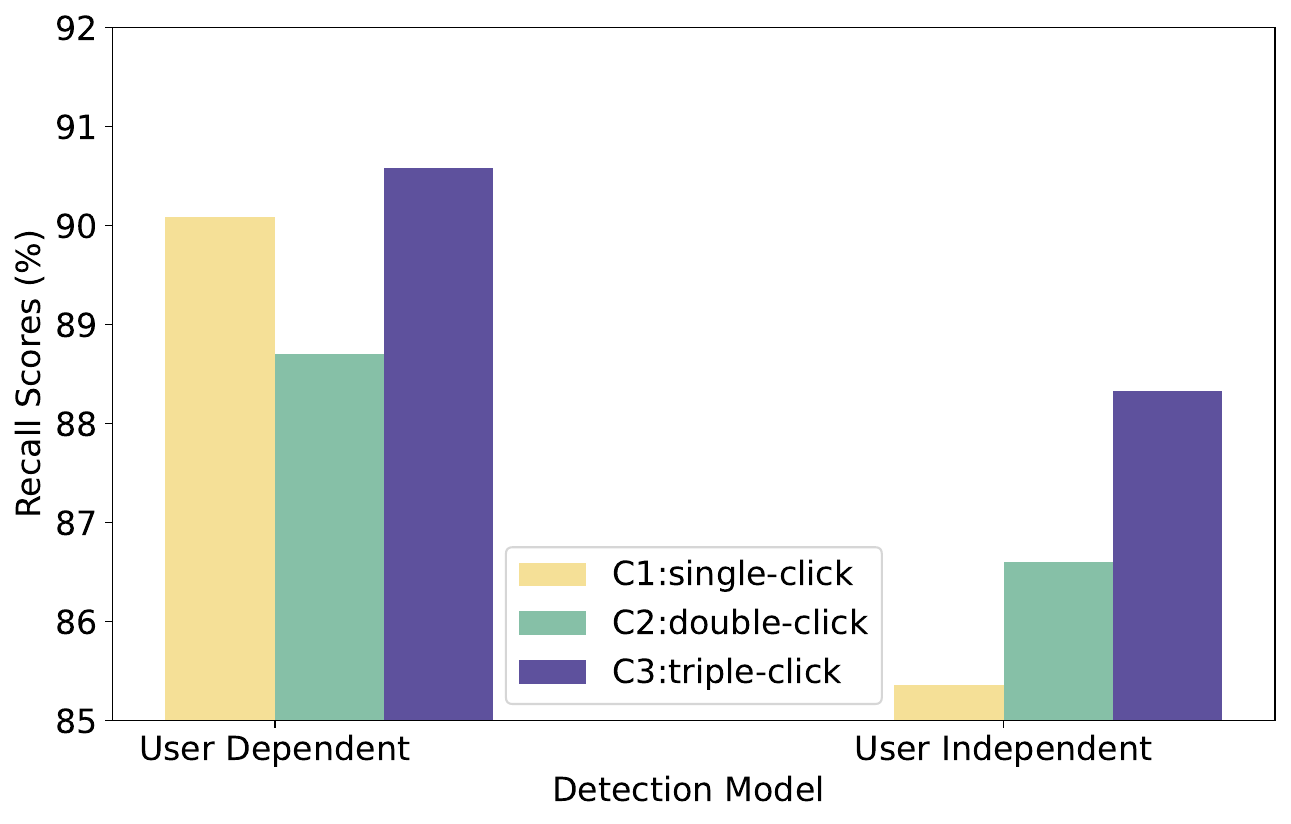}{fig:s2-res-recall}{\change{Recall (\%).}}{0.49}{graph}

\caption{Comparison of the accuracy in the two detection models across the three gestures. Both models have good detection accuracy but the user-dependent model outperforms in single and triple clicks.}
\label{fig:s2-res-accuracy}
\end{figure}

\ressec{Gestures' Performance Time} The median time for performing the \cone is 3.53 seconds (min= 0.5 and max= 7.17), followed by 4.47 seconds for \ctwo (min= 2.65 and max= 6.3), and 5.32 seconds for \cthree (min=3.8 and max= 6.25). 


\subsubsection{RQ1: Which Gesture is More Accurate to Detect?}
\missingpax{22}{4} Figure \ref{fig:s2-res-accuracy} summarizes the results. We wanted to understand three aspects:
\begin{enumerate}
\item For each gesture separately, what is the most accurate detection model?
\item For each model, is there a favoured gesture in terms of detection accuracy?
\item Assuming we selected the best detection model for each gesture, which gesture is most accurately detected?
\end{enumerate}
\changedelete{To answer Point 1, we separately compared for each gesture the model type for detection (user-dependent and user-independent model) to the accuracy (precision and recall)}\change{To address Point 1, we compared the two detection model types (user-dependent and user-independent) separately for each gesture, evaluating their accuracy in terms of precision and recall}. Looking at \cone, both models performed well with mean precision equal to 90.7\% for the user-dependent model and 85.4\% for the user-independent model. The data shows strong evidence that the user-dependent model outperforms the user-independent model and yields higher accuracy \bfval{43.028}. However, the data is inconclusive whether this is in precision or recall only or both though the trend shows ca. 5\% more in both precision and recall \bfval{0.431}. Similarly, there is moderate evidence for \cthree that both models perform well (mean precision: user-dependent = 91.1\% and user-independent= 88.9\%), with the user-dependent one outperforming by approximately 2\% \bfval{3.695}. Similar to \cone, the data is inconclusive whether this is in precision or recall only or both \bfval{1.354}. For \ctwo, the data is inconclusive if there is a difference in detection power between the models or not, with the data trend indicating both are equally good \bfval{0.451}. Synthesizing the three findings, this indicates that both models are good to use with the user-dependent one outperforming the user-independent model.  

To answer Point 2, we separately compared for each model, the accuracy (recognition and recall) for each of the gestures (\cone,\ctwo, and ~\cthree). The results were inconclusive with a trend towards no difference in accuracy detection between the gestures in the two model types (user-dependent \bfval{0.959} and user-independent \bfval{0.447}). 
However, we observed moderate evidence that the precision is significantly higher than recall using the user-dependent model in \ctwo by ca. 1\% \bfval{4.769}. This finding is irrelevant to our analysis and thus we do not follow up on it. 

To answer Point 3, we chose to compare the accuracy using the \emph{user-dependent model} for the detection of the three gestures based on the results of Point 1 (generates significantly higher accuracy for \cone and \cthree + seems plausible for \ctwo). We did not re-perform the test as it corresponds to the results in Point 2 because we did not need different models across the gestures. 

\takeaway The user-dependent and user-independent models both are accurate for detecting the gestures with relatively high precision and recall \mrngbig{85\%}. However, the \emph{user-dependent} model outperforms the user-independent one specifically with \cone and \cthree. The expected gains are in the order of ca. 5\%. 



\insertfig{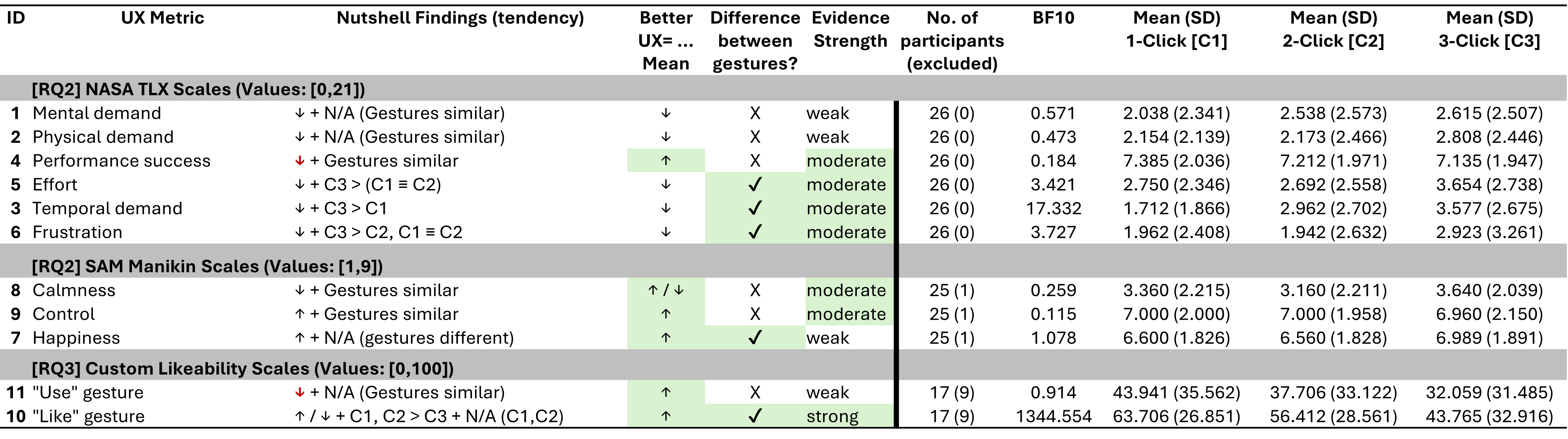}{fig:s2-res-tbl-ux}{Summary of the questionnaire's results examining the user experience while doing the three gestures. All gestures were not frustrating and easy to perform (evident in the low means, scales in Section \ref{sec:s2:analysis}).Users preferred the single click and disliked the triple click. $\uparrow$ denotes higher values  and $\uparrow$ denotes lower values. ``Better UX'' column tells us whether higher or lower values indicate a better user experience. The arrows in the ``Nutshell findings'' column tells us whether the mean values are high or low compared to the center of the scale.}{1}{table showing the bf values}

\insertfig{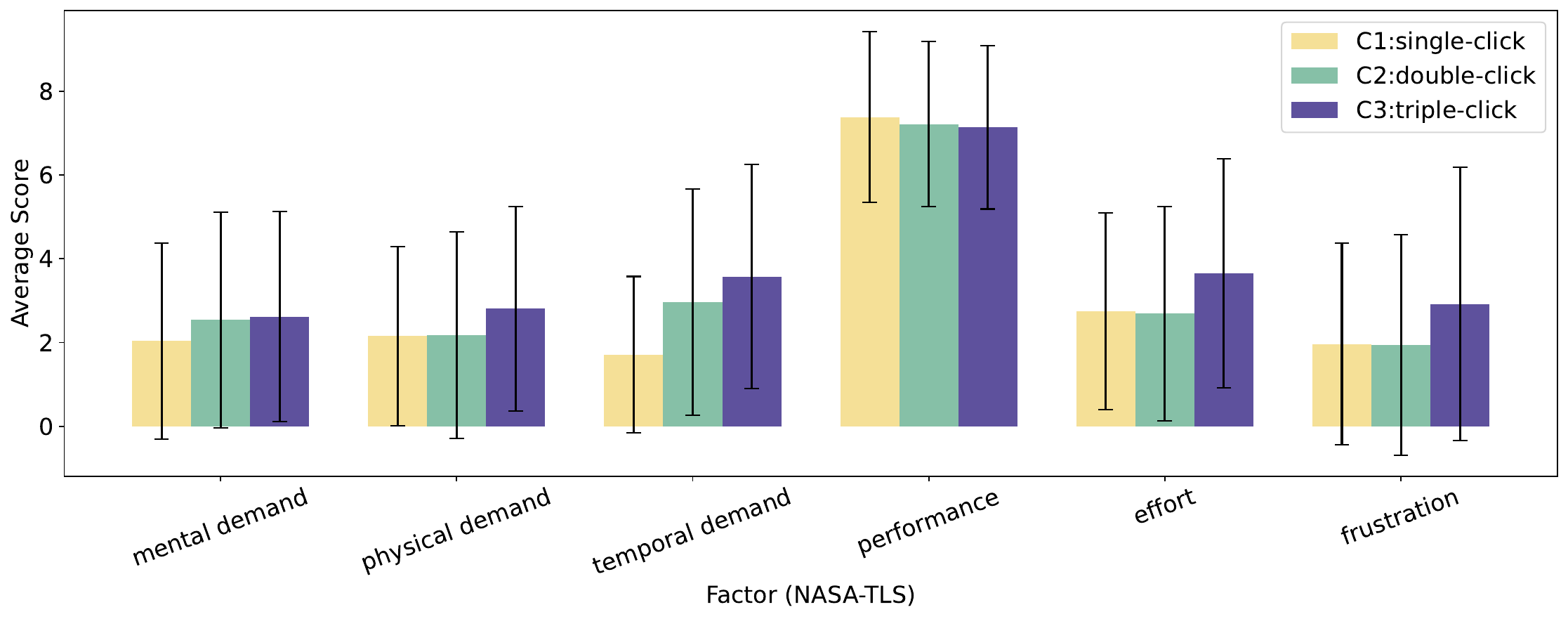}{fig:s2-res-nasa}{Comparison of the three gestures using NASA-TLX. The maximum score is 21. }{1}{NASA TLX graph}

\begin{figure}

\insertsubfig{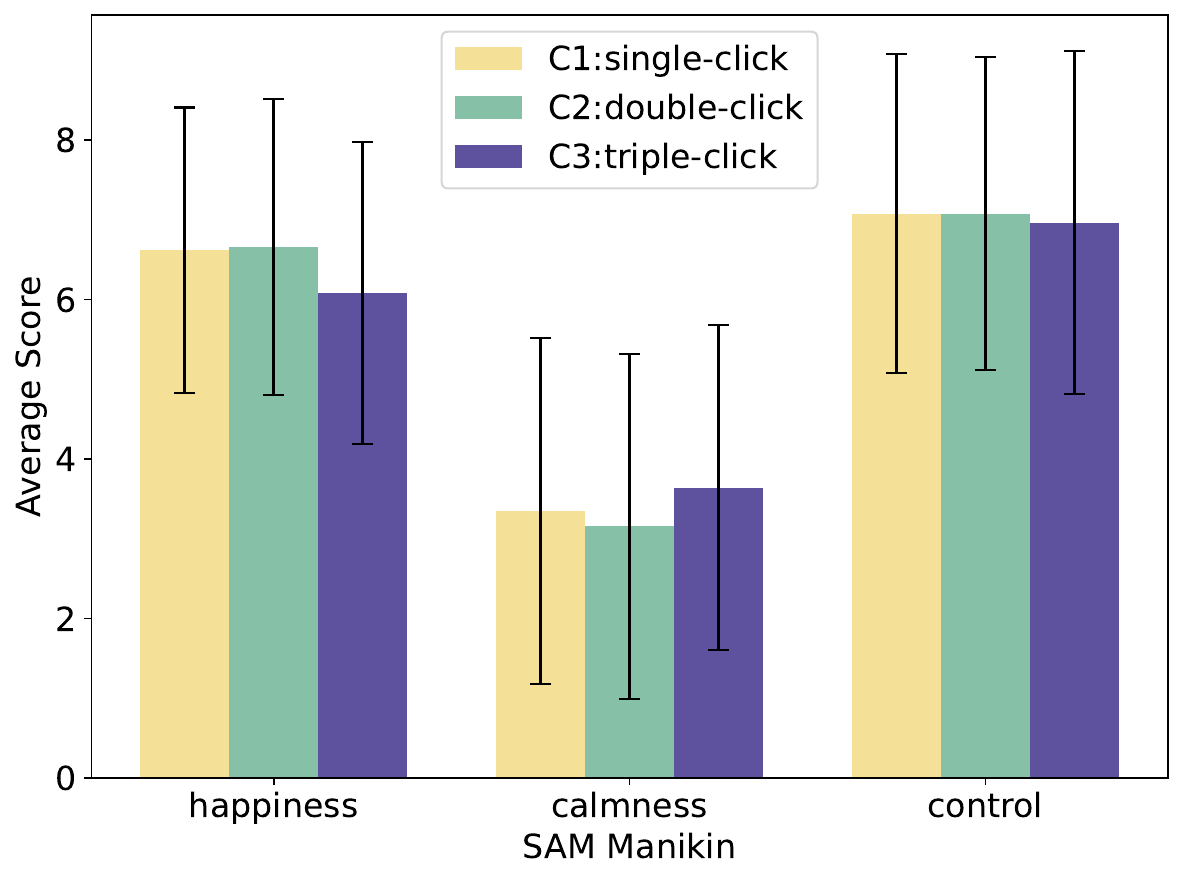}{fig:s2-res-sam}{SAM Manikin.}{0.48}{graph}
\hfill
\insertsubfig{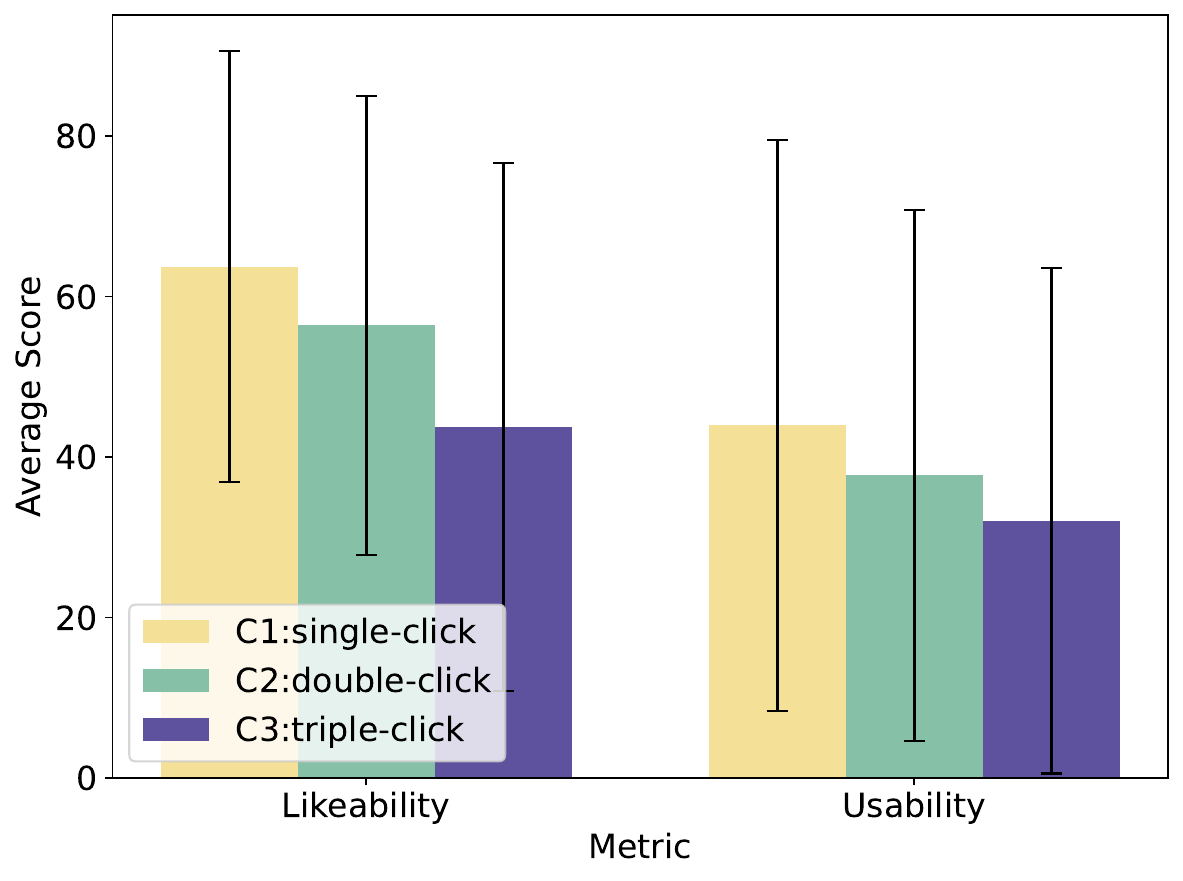}{fig:s2-res-custom}{Likeability and Usability Custom Scales}{0.48}{graph}

\caption{Comparison of the users' feelings and preferences about the three gestures. Participants were happy, calm and in control regardless of the gesture type. They favoured single clicks the most and disliked triple clicks. In general, they did not think they would use the gestures in daily life.}
\label{fig:s2-res-feelings}
\end{figure}

\subsubsection{RQ2: What was the Users' Experience with the Gestures?}

 Table \ref{fig:s2-res-tbl-ux} summarizes all the statistical values and findings from the questionnaire (see Appendix \ref{sec: Questionnaire}). We start by reporting on the results of the NASA-TLX (summarized in Figure \ref{fig:s2-res-nasa}). We analyzed the data from the 26 participants. All gestures were not demanding mentally \mrng{2.6} or physically \mrng{2.8}, suggesting they are easily executable. The data trend suggests there was no difference between the gestures in physical or mental demand, but the results are inconclusive. This aligns with the next data point, where perceived effort to perform all gestures was generally low \mrng{3.6}. However, there is moderate evidence that participants perceived \cthree to require significantly more effort than \cone and \ctwo ($BF_{10}= 3.092 ~and~ 3.105~ respectively$), while there was no difference in the perceived effort between \cone and \ctwo \bfval{0.210}. The three gestures had low temporal demand and the participants did not feel rushed performing them \mrng{3.6}. However, there is strong evidence that participants felt more rushed performing \cthree compared to \cone \bfval{41.054}. The difference between \ctwo and (\cone or \cthree) is inconclusive, though the trend suggests no difference between \ctwo and \cthree \bfval{0.43} and participants being more rushed in \ctwo than \cone \bfval{1.69}. Despite the low effort, there is moderate evidence that participants did not feel they performed the gestures successfully, regardless of the gesture type \mrngbig{7.1}.  Unexpectedly, despite the low perceived success, participants generally reported low levels of frustration across all gestures, with a rating of \mrng{2.9}. However, there is moderate evidence that participants felt more frustrated performing \cthree compared to \ctwo \bfval{3.262} and no difference in frustration between \cone and \ctwo \bfval{0.208}. The data trend suggests \cthree is also more frustrating than \cone \bfval{2.578}, but the results are inconclusive.  

Next, we discuss the results from the SAM-Manikin to understand the participants' emotional experiences during the gesture execution rather than their subjective performance. \missingpax{25}{1} Figure \ref{fig:s2-res-sam} visualizes the results. There is moderate evidence that participants felt generally calm \mrngbig{3.2} and in control \mrngbig{7} during gesture execution, regardless of the gesture type.  They were also happy while performing the gestures \mrngbig{6}. However, we cannot make claims about whether the gesture type affects their happiness score as the data is inconclusive. 

\takeaway Overall, all gestures were positively perceived by the participants and have a \emph{good user experience}. They were not frustrating and easy to perform (low effort, mental, physical, and temporal demand). However, the participants did not think they were executing them successfully. Participants felt calm, in control and happy while performing all of them. The SAM-Manikin did not indicate a preference towards one of the gestures. However, the NASA-TLX indicated that \cthree was the least favourite (higher: temporal demand, perceived effort, and frustration) and \emph{\cone was the most favourite}. The data trends of \ctwo indicate potential comparability to \cone but the current results are inconclusive. The preference for \cone is positively surprising, as we anticipated participants would dislike it for false positives (interference with regular breathing), and \cthree would induce exhaustion due to its duration. Consequently, we expected a distinct preference for \ctwo. However, the data refuted those assumptions.


\subsubsection{RQ3: Which Gesture did the Users Prefer?}
\missingpax{17}{9} Table \ref{fig:s2-res-tbl-ux} summarizes all the statistical values and findings from the questionnaire (see Appendix \ref{sec: Questionnaire}). Strong evidence suggests a variation in participants' perceptions of gesture likeability. Participants liked \cone and \ctwo and disliked \cthree. The posthoc test indicates that users preferred \cone \mean{63.7\%} and \ctwo \mean{56.4\%} compared to \cthree \mean{43.8\%} (\cone= strong evidence \bfval{1470.003} and \ctwo= moderate evidence  \bfval{9.984}). The data suggests a distinction in user preferences between \cone and \ctwo but the results are inconclusive  \bfval{1.631}. Conversely, users showed little interest in utilizing the proposed gestures for interaction (all mean scores below 44\%). The data also suggests no difference between the users' desire to use the three gestures but the results are inconclusive. Looking at the individual gestures, 47\% of the participants wanted to use \ctwo, 41\% for \cone, and 29.4\% for \cthree.

\takeaway While the majority of users were not interested in using gestures for interactions such as controlling a device in the room or doing a task on the PC/mobile, a decent portion (over 41\%) was interested in using double click and single click. From a likability perspective, users preferred the single-click and the double-click gestures to the triple-click gesture. 


\subsubsection{Qualitative Reflections}
We reflect here on the results of the post-study interview. Participants’ post-study sentiments were diverse: 8 felt positive or neutral, 7 felt fatigued or breathless, and 6 felt unfamiliar, self-conscious, or bored. The main challenges included concerns about accidental activation and session length, noted by 8 participants. Three participants thought the gestures interfered with their natural ability to breathe and speak. This provides insight into the threshold for participant fatigue, based on our use of 60 trials.

Most participants (15 out of 26) found the visualization easy to grasp, with 19 quickly memorizing gestures after 5 trials lasting about 50 seconds. The majority (20 out of 26) agreed the training was sufficient and intuitive without extra explanation. The majority also agreed that the training time was sufficient (n=18).

Opinions on using breathing-clicking gestures for button activation were divided: 10 were in favour, 8 were against, and 1 was undecided. Similarly, 10 participants agreed to use gestures for environmental interaction, while 4 declined. Most were satisfied (14) or neutral (6) about the interaction technique. Common application areas included "turning objects on/off" (14 mentions), particularly lights\footnote{There was a single mention to TV, doors, or fans.}, and "special user groups" (10 mentions), notably persons with disabilities (6 mentions). Other suggestions included older adults, breath monitoring for relaxation, and minimal mentions of virtual reality games, emergencies, and button activation. The identified application areas underscore the need for focused research, given the novelty of the technology and participants' unfamiliarity with its potential applications.

\subsection{Takeaways}
In a nutshell, we most accurate detection model is the \emph{user-dependent} one and the favourite gesture is the \emph{single-click} gesture, while the least favourite is the \emph{three-click} gesture. We enlist below the key findings used to derive the design guidelines in the next section. 
\begin{enumerate}
\item User-dependent models for detection outperform the user-independent ones with an accuracy exceeding 90\% and a training set of 14 instances conducted in approximately one minute. However, user-independent models present a viable alternative, providing relatively high accuracy (above 85\%) when user calibration is impractical (answer to RQ1).
\item The user experience with performing the gestures was positive and they are all usable. The gestures were easy to perform and users felt good while performing them. Interestingly, participants did not subjectively feel they were performing the gestures correctly although we can see from RQ1 that the detection accuracy is high. Combining the various UX metrics and the participants’ explicit interest, \cthree was the least favourite and \cone was the favourite (answer to RQ2). 
\item Despite the positive user experience, participants were not interested in using gestures for interaction. It is not clear whether this reflects a genuine desire or a ``novelty effect’’ situation where they could not imagine how the interaction would look like (answer to RQ3). \todofu{reflect on qual data}
\item The data about \ctwo is generally inconclusive in terms of the best models for detection, the user experience and the user preferences. The data trends suggests it is similar to \cone in terms of  user-experience and its detection accuracy is high with both models.   
\end{enumerate}

\section{Part 3: Design Guidelines, Reflections and Limitations}
\label{sec: discussion}


\insertfig{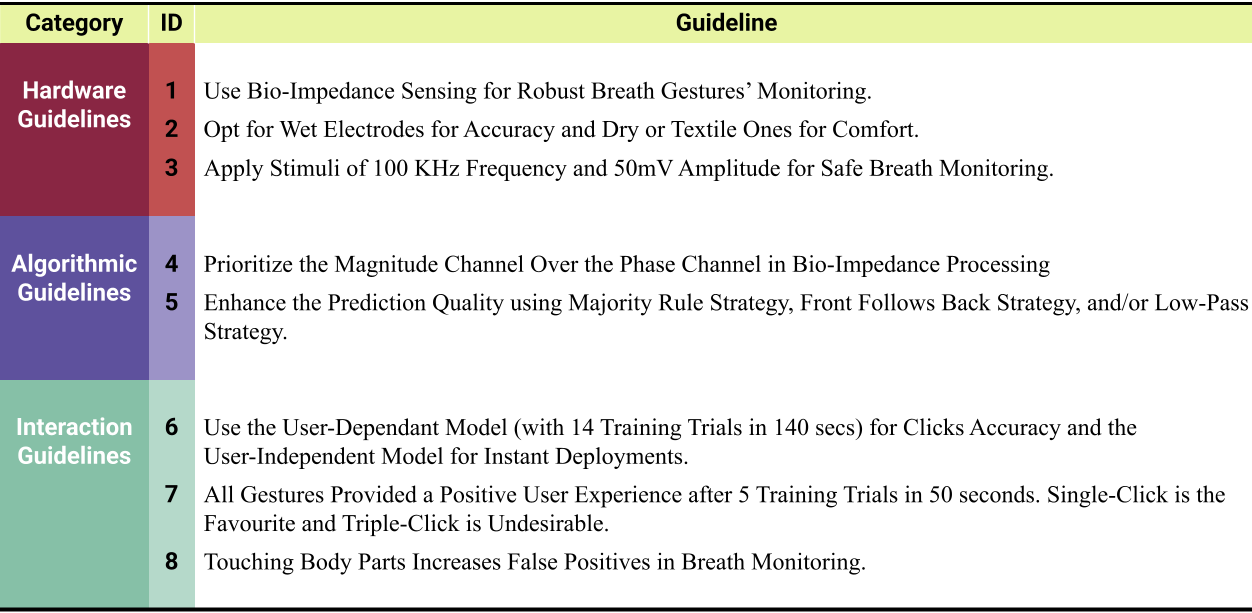}{fig:guidelines}{Guidelines for designing breathing gestures synthesizing the findings from both studies. For details, see Section \ref{sec: discussion}.}{1}{guidelines table showing 8 of them}

\begin{table}
\centering
\caption{Comparison between the average time for a 1-click using different actuators. While the breathing click ``iBreath'' is slower than traditional mouse clicks, this is expected for hands-free interactions, which complement rather than replace traditional clicks.}
\begin{tabular}{|l c|} \hline  
\textbf{Input Method} & \textbf{Time (ms)} \\ \hline  
 \multicolumn{2}{|c|}{\textbf{Hand Clicks}}\\ \hline  
Mouse Click & 150–250 \\  
Keyboard Click & 150–200 \\  
Touchpad Click & 180–250 \\ 
Hand Gesture Click & 300–600 \\ 
Air Tap Gesture & 500–1000 \\  \hline 
 \multicolumn{2}{|c|}{\textbf{Hand-Free Clicks}}\\ \hline 
Foot Pedal Click & 150–250 \\
Eye Gesture (Blink) & 200–500 \\ 
Eye Gesture (Dwell) & 400–800 \\  
Voice-Based Command & 500–1000 \\   
Head Movement Click & 500–1000 \\ 
EEG-Based Click & 1000–1500 \\   
\textbf{Breath Click ``ibreath''} & 3500 \\ \hline 

\end{tabular}
\label{tbl:clickspeed}

\end{table}


 \subsection{Design Guidelines}
We present below three guidelines for designing the hardware (\#1 $\rightarrow$ \#3), two guidelines for designing the algorithm (\#4 and \#5), and three guidelines for designing the interaction (\#6 $\rightarrow$ \#8) as shown in Figure \ref{fig:guidelines}.

\guideline {Use Bio-Impedance Sensing for Robust Breath Monitoring}
Bio-impedance sensing-based breath monitoring solution is proved in this work that it can provide robust measurement results with a concise hardware system based on the AFE chip AD5941 and the reported detection algorithms (user-dependent or user-independent) in the paper. 

\guideline {Opt for Wet Electrodes for Accuracy and Dry or Textile Ones for Comfort}
The electrode plays an important part in bio-impedance sensing-based breath activity monitor systems. While wet electrodes offer high-quality signals, they may result in discomfort during removal due to the electrode glue. Our interview findings echoed that concern about the obtrusiveness and the portability of the hardware. Other electrodes such as dry electrodes or textile electrodes could provide a more comfortable user experience than wet electrodes. However, the signal quality could be reduced. 

\guideline {Apply Stimuli of 100 KHz Frequency and 50mV Amplitude for Safe Breath Monitoring}
The voltage stimuli with 100 kHz frequency and 50 mV peak-to-peak amplitude can ensure that the amount of current in the body is under the maximal safety current of 0.6 mA and provide a good-quality signal. This is validated in our findings from Part 1 and 2, where there is no registered uncomfortable experience during both studies.


\guideline{Prioritize the Magnitude Channel Over the Phase Channel in Bio-Impedance Processing}
The bio-impedance signal's distinct features during breath gestures are efficiently handled by a lightweight neural network, alleviating participant concerns about confusion with regular breathing. In bio-impedance sensing, the magnitude channel is favoured over the phase channel for a clearer signal. This should be reflected in their respective weights when designing a sensor fusion algorithm. It is important to recognize that using more input channels increases the number of model parameters and the model size. This can be problematic for edge devices with limited hardware resources. If users must choose only one channel due to hardware constraints, prioritize the Magnitude channel. 

\guideline {Enhance the Prediction Quality using Majority Rule Strategy, Front Follows Back Strategy, and Low-Pass Strategy}
We recommend that designers identify two aspects about any breathing gesture: it’s expected execution duration and the characteristics of the breathing patterns involved.  Understanding the anticipated duration of the gesture can significantly enhance performance by informing the sizing of the sliding window. We recommend three strategies for improving the performance based on the gesture’s pattern. Use the ``majority rule’’ strategy when several continuous predictions cover an instance of a gesture. An example is a gesture that lasts for one second and the sliding window is every 5 milliseconds resulting in 20 prediction points per instance. Use the ``front follows back'' to mitigate model confusion from recognizing two or more gestures with overlapping mini patterns. An example is a double-click could be mispredicted as two consecutive single-clicks.  Use the ``low-pass’’ strategy to reduce singular false positives resulting from idle behaviour. An example is if a user is inactive and simply moving around with their hands touching, altering the impedance and resulting in a false positive gesture. See Section \ref{sec:comp3algo} for technical details.


\guideline {Use the User-Dependent Model (with 14 Training Trials in 2.3 Minutes) for Clicks Accuracy and the User-Independent Model for Instant Deployments}
Both detection models performed strongly in terms of prediction accuracy and completeness of click gestures (precision and recall). However, the user-dependent model exhibited superior accuracy (over 90\%) with just 14 calibration points per gesture, lasting approximately 140 seconds. This model is well-suited for systems involving prolonged user interactions and/or sensitive actions, such as home deployments, where minimal training is desirable compared to extended usage times and potential frustration resulting from lower accuracy.
On the other hand, the user-independent model also yielded satisfactory results (over 85\%) after pre-training on the data of 21 participants. Thus, designers and researchers can employ it directly as a plug-and-play solution without the need for custom model training. This model is more suitable for immediate interactions where system usage is limited, the performed action is not critical or calibration is not feasible (e.g. in public spaces). 

\guideline {All Gestures Provided a Positive User Experience after 5 Training Trials in 50 seconds. Single-Click is the Favourite and Triple-Click is Undesirable.}
Using subjective quantitative metrics (NASA-TLX, SAM Manikin, and Likeability scales), participants found all gestures usable, easy to perform and felt calm, happy and in control while performing them. However, they exhibited a clear preference for single clicks and disliked triple clicks. The data regarding the double-clicks was inconclusive although the trend suggests it is comparable to single-clicks. 

Participants mentioned in interviews that the training period was adequate and found the ball-graph visualization (see Figure \ref{fig:s2-method}) self-explanatory for learning new gestures. We suggest instructing users to perform any new gesture five times while observing a graph that depicts the gesture. The technique directs them to breathe following the ball where graph peaks represent inhalation and troughs represent exhalation. The training session takes 50 seconds (10 seconds per instance). \todofu{add qual data}

\guideline {Touching Body Parts Increases False Positives in Breath Monitoring}
Touching a body part with the hand creates an alternative electric current path to the one passing through the lung and is measured here using bio-impedance sensing to monitor breathing. This phenomenon introduces dynamic noise, complicating the detection of the breath gesture. This poses an interaction design challenge to use the gestures reliably. To overcome this, future work can explore two approaches: introducing self-touching activities to the training set or using the data augmentation method to simulate such disturbance and improve the robustness of the neural network against such noise.

 \subsection{Application Areas}
 
 Breath-clicks (takes 3.5-5.3 seconds\footnote{This is the average duration of all gesture types: single-click, double-click, and triple click not only the single click}) are significantly slower than traditional mouse clicks, eye-clicks, and voice-clicks, making them unsuitable for rapid consecutive actions like office tasks as opposed mouse clicks. \cref{tbl:clickspeed} contextualizes our results showing that while the breathing clicks are slower compared to mouse clicks, this is expected for hand-free clicks. However, 
 \changedelete{breath-clicks excel in specific scenarios: 1) when hands and eyes are unavailable for interaction, such as in surgical settings; 2) controlling smart environments (e.g., turning lights on/off or adjusting volume) without needing to see the device or press a button; 3) controlling wearable devices with limited surface area for touch interaction (e.g. volume control for airpods headphones), 4) triggering actions invisibly or discreetly, and 5) aiding patients with severe disabilities or those intubated in hospitals to perform actions like calling a nurse. We acknowledge that skin-attached electrodes in the current setup are impractical for mobile wearables (though possible). Future developments could integrate electrodes into undergarments and use the same algorithm on mobile phones to process the gestures or explore bio-impedance sensing in alternative areas, such as the ears, using sensor-equipped headphones.}
 \change{breathing-based control offers distinct benefits over voice, vision, or IMU-based hands-free inputs, especially in contexts where those modalities are impractical. Voice interfaces, while powerful, are not universally usable: they fail when users cannot speak or when speech would violate privacy or silence requirements (e.g. in a library or during a meeting). Vision-based gesture systems require line-of-sight and favorable lighting, and can be socially intrusive or unreliable under occlusions. IMU-based inputs (e.g., head movements or wearables detecting motion) demand deliberate physical actions, which may be fatiguing, conspicuous, or impossible for users with motor impairments. In contrast, breathing gestures are \textbf{silent, subtle, and accessible}. Prior research has noted that traditional modalities like touch, voice and hand gestures are “not inclusive of the entire population” and cannot be used by people with impairments in those channels \cite{hundia2019breathin}.
Breathing, however, ''can be issued by anyone who can breathe'', making it a truly ubiquitous capability independent of speech or limb motion. Moreover, breathing gestures can be performed with minimal external visibility or noise, users can modulate their breath without drawing attention. This discretion is a unique strength: breathing acts as an invisible interface that bystanders are unlikely to notice, whereas speaking aloud or broad gestures readily expose the interaction.
In summary, iBreath’s breathing-based control is advantageous whenever audible or visible inputs are unsuitable. It combines the hands-free convenience of voice or gesture with a level of stealth and inclusivity those modalities cannot match. This makes breathing control especially attractive for assistive technology and situations requiring subtle, eyes-free interaction.}

\change{Beyond previously discussed applications, additional compelling use cases include: (1) immersive experiences, such as gaming or virtual reality, where breathing gestures can serve as intuitive, natural interaction inputs enhancing immersion without disrupting the experience; (2) mental health and stress management applications, where breathing patterns can trigger therapeutic interventions or mindfulness prompts unobtrusively; (3) covert security scenarios, allowing personnel to silently and discreetly communicate or trigger alerts without visible movements or sounds; and (4) augmented reality (AR) and wearable computing, enabling subtle interactions (e.g., navigating menus, selecting items, or toggling visual overlays) without drawing attention in social or professional contexts.}

 \subsection{\change{Long-Term Viability and Adoption Potential}}
\change{For breathing gestures to become a sustainable input method, they must be \textbf{easy to learn, socially acceptable, and compatible with everyday devices}. Our findings and prior studies (e.g., Burr et al., 2023 \cite{burr2023breathtures}) confirm that users can quickly learn and reliably perform breathing gestures, even if some patterns (like sharp inhales) require brief practice. Participants in our study found the gestures natural after minimal exposure, indicating strong learnability.
Breathing gestures also excel in social acceptability. Unlike voice or large gestures, they are subtle and easily concealed within natural actions like sighs or deep breaths, avoiding the awkwardness often reported with public interactions. Our users expressed no discomfort and valued the discreetness, suggesting good prospects for wider acceptance as familiarity grows.
Technologically, breathing gestures are well-suited to integration with wearables. 
While we acknowledge that using the skin-attached electrodes for sensing in the current setup is inconvenient for mobile wearable scenarios, one could imagine further developing the prototype in the future and integrating the dry electrodes into undergarments.
For example, bio-impedance or stretch sensors can be embedded into straps or clothing with integrated textile electrodes. As a passive, low-power channel, breathing input can run continuously in the background, enabling seamless, eyes-free control alongside daily activities.
In the long term, breathing gestures could complement other modalities (e.g., gaze for pointing, breath for selection) to support rich multi-modal interactions. With minimal hardware demands and natural usability, iBreath offers a practical, scalable pathway for breathing-based interaction in future wearable ecosystems.}

\subsection{\change{Robustness and Stability of Bio-Impedance Sensing}}

\change{A key consideration for iBreath’s long-term deployment is the robustness of the bio-impedance respiratory sensing and the system’s reliability under real-world conditions. Bio-impedance has the advantage of being a direct, contact measurement of breathing (sensitive to chest expansion and contraction), but like any physiological sensor, it can be influenced by various factors. We here discuss potential issues, including physiological changes (fatigue, stress), environmental conditions (sweat, temperature), and user motion, and how our system can mitigate them.}

\subsubsection{\change{Physiological Variability}}
\change{Breathing patterns naturally vary with fatigue, emotional states, or illness, potentially affecting gesture recognition. For instance, heavy breathing from exercise might resemble intentional gestures, while shallow breathing during stress could complicate gesture detection. To address this, iBreath employs the data augmentation method to simulate such noise and improve the robustness of the model during training. 
Besides, gesture detection relies on relative changes in impedance rather than absolute values, enhancing resilience against variations in baseline breathing intensity. Prior studies support this adaptive approach; Goyal et al. demonstrated that although absolute bio-impedance measurements can vary due to skin hydration or electrode placement, the relative breath-induced impedance changes remain consistent and reliable \cite{goyal2024day}. Additionally, implementing drift compensation, periodically resetting baseline impedance, ensures stable and robust gesture recognition despite physiological and environmental fluctuations.}

\subsubsection{\change{Environmental and sensor factors}}
\change{Sweat and motion artifacts commonly challenge wearable sensors. Sweat might alter electrode-skin impedance or affect electrode adhesion; however, bio-impedance respiration sensors inherently tolerate moderate variations in skin moisture. Prior research indicates that skin hydration does not significantly affect respiratory impedance measurements in a relative sense \cite{goyal2024day}. 
Besides, in the future, it can utilize moisture-resistant textile electrodes to maintain signal integrity during perspiration.
 Extreme conditions (soaking wet or very dry skin) might introduce some noise, but those can be addressed by using adaptive filtering or even sweat-resistant electrode materials (e.g., hydrophobic coatings) as demonstrated in emerging wearable designs. Motion artifacts are another challenge: when the user moves vigorously (e.g., running or jumping), the chest strap sensor may shift or experience pressure changes unrelated to breathing. To improve robustness, the system can leverage multi-modal sensing. For example, an accelerometer can be integrated to distinguish deliberate breathing gestures from general body motion. If a large movement is detected coinciding with a potential breath gesture, the system can require a higher confidence (or an additional sensor confirmation) before registering it as a command. 
 Such redundancy ensures reliability even if one sensor modality is momentarily compromised.}

\subsubsection{\change{Long-term stability}}
\change{For bio-impedance sensors, long-term use (months of daily wear) highlights the importance of sensor calibration and consistent placement. Prior studies demonstrate that well-calibrated bio-impedance systems reliably track respiration over extended periods, correlating strongly with actual lung volumes \cite{goyal2024day}. 
However, daily variability in sensor placement or physiological changes can significantly affect raw impedance measurements, emphasizing the necessity of routine calibration and consistent positioning. To address this, iBreath can integrate quick recalibration routines each time the device is donned, ensuring readings remain aligned to a standardized baseline in the future. Furthermore, by continually updating a user-specific profile through machine learning, the system becomes progressively tailored to individual respiratory patterns, improving accuracy over time. Gesture designs leverage pronounced impedance changes (such as strong inhales or extended exhales), minimizing the likelihood of misclassification from minor day-to-day respiratory fluctuations due to fatigue or emotional states.}

\subsection{Limitations and Future Work} The skin attachment of electrodes may become uncomfortable over prolonged periods. While complaints from participants were minimal, future research should focus on designing more comfortable electrodes. Two limitations in interpreting our results include the lack of real-time gesture detection and participants avoiding interference with impedance by not touching other body parts. 
Moreover, the robustness of the breathing gesture detection system was only evaluated in three scenarios (walking, lying and sitting). We recommend an in-the-wild experiment in the future, particularly in breathing-intensive scenarios that could interfere with the gestures such as doing sports. 
Our future studies will focus on real-world interactions, challenges, and applications, expanding beyond system creation and gesture design (the primary focus of this paper).Further research should also explore the threshold for reducing the current interaction time (3.5-5.3 seconds per gesture in intervals of 10 seconds) without compromising participant comfort. Similarly, future research could look into empirically reducing the training time (e.g. in a subsequent pilot study with three participants, two training instances were enough for participants to learn the gesture).  Further research is needed to test physique's impact on gesture detection. For example, the impact of body fat on impedance changes was not explored, potentially limiting viable user groups. Additionally, the effect of respiratory-related parameters, like lung volume, respiratory frequency, respiratory rate, and dominant breathing pattern on the system's performance could be explored.  A common limitation, which we also face, is the limited age range of participants, presenting an opportunity to explore the system's acceptability with diverse user groups in the future. Lastly, while our system may benefit both able-bodied and differently-abled people, our testing pool only included able-bodied people. Future studies should involve differently-abled people, who may be the primary beneficiaries of our approach. 

\section{Conclusion}
\label{sec: conclusion}
This paper presents the design of a new system ``iBreath'' to detect breathing gestures. Our system uses low-cost wearable hardware setup detecting bio-impedance variation caused by breath activity and an optimized proprietary detection algorithm that we designed. The system was tested through two lab user studies (total n= 34) to understand the robustness of the designed gestures detection, the users' comfort, and their general experience while using them. We mainly tested three gestures mimicking famous button clicks: single-click, double-click, and triple-clicks. 

Our results show that the system successfully detects breathing patterns and more specifically gestures. The three designed gestures are easy and comfortable to use. We also show that it is feasible to train users on entirely new gestures after performing them only five times (in approximately 50 seconds) with the help of our training visualization. The users preferred the single-click gesture and disliked the triple-click gesture. We also show that training the detection models using only the user's data (user-dependent) or collective data from others (user-independent) both offer high accuracy in predicting the gestures. However, the user-dependent is superior by approximately 5\% once the user performs the gesture 14 times for training (takes 140 seconds).  


Although breath rate monitoring using bio-impedance has been extensively studied, the detection of breath gestures using this sensing method remains under-explored, representing the novelty of our work. Our work offers a toolkit for designers to build new breathing gestures, enables designers to directly explore clicking interactions based on our proposed gestures, and provides developers with guidelines on how to extend the detection to other gestures. The portability of the wearable sensing electrode remains an intriguing research challenge warranting further investigation in the future.


\begin{acks}
The research reported in this paper was supported by the European Union’s Horizon Europe research and innovation programme in the Project STELEC under grant agreement No 101162257.
We want to acknowledge that ChatGPT-4.0 which has been used to restructure some sentences in this manuscript following current ACM policies.
\end{acks}
\bibliographystyle{ACM-Reference-Format}
\bibliography{papers}

\newpage
\appendix


\section{System Design: Hardware Design} \label{appendix:hw}
\insertappendix{System Design: Circuit Schematic}{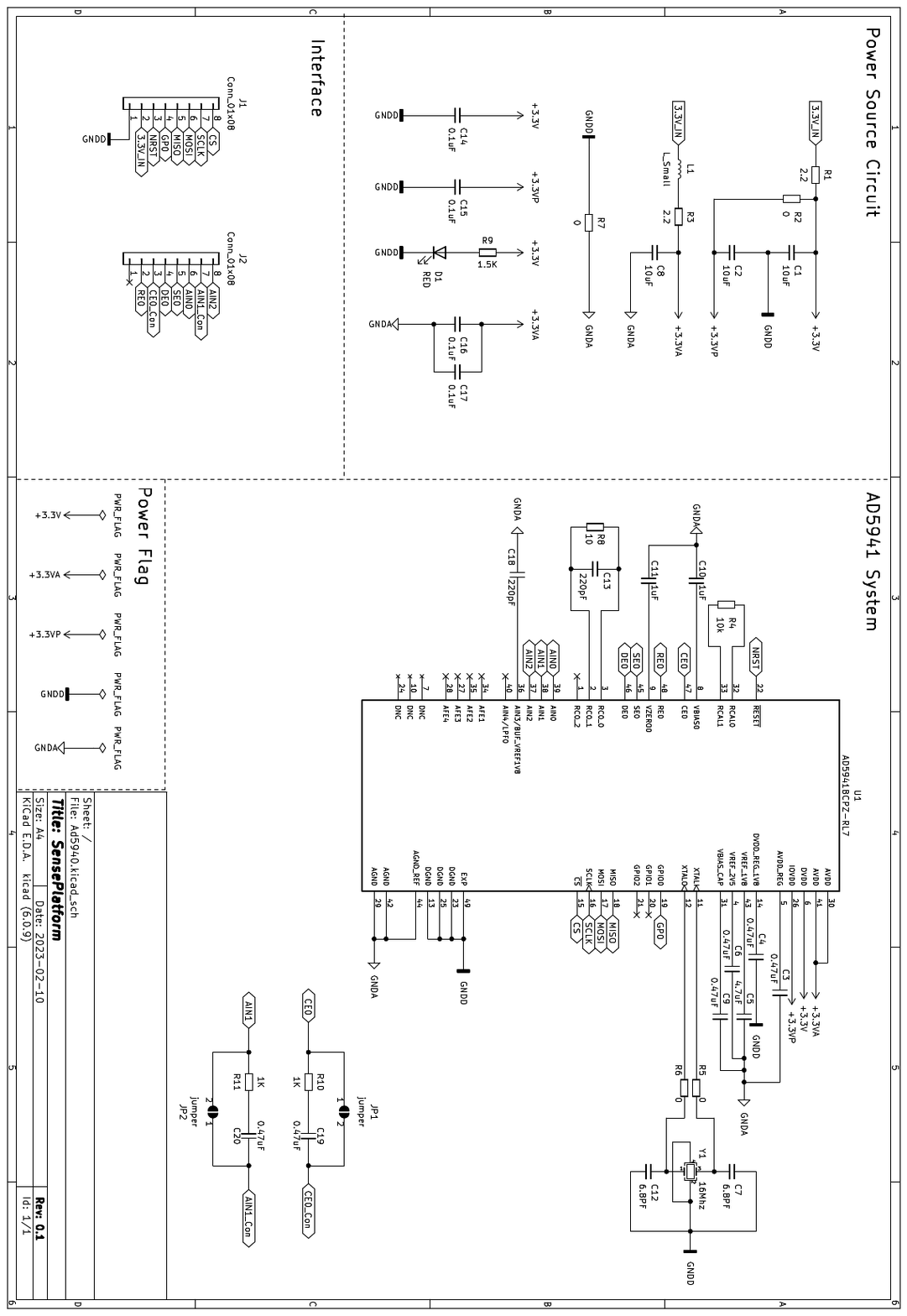}{sec: circuit}
\subsection{Technical Details about the Sensor Design}
Bio-impedance, unlike time domain signal measurement, belongs to the frequency domain and thus demands a more complex setup. This involves using an alternating voltage or current as the stimulus and a Fourier Transformation engine to extract frequency domain information from both the response and stimuli signals.

The system has 2 main components: the Analog Front-End (AFE) and the microcontroller. The AFE employs the integrated chip AD5941 from Analog Devices, capable of generating an alternating voltage ($V(w)$) within a frequency range of 0.015 Hz to 200 kHz, and measuring the alternating current ($I(w)$), via the Trans-Impedance Amplifier. This chip is also equipped with a Fast Fourier Transformation (FFT) hardware accelerator, enabling rapid acquisition of frequency information. To account for the latency between stimulus generation and response measurement, which can introduce measurement bias, the iBreath employs a ratio measurement method. This involves measuring the bio-impedance of the object under test by first measuring the current through a high-precision known resistor ($Z_{known}$) as ($I_{known}$), and then applying the same stimulus to the test object and measuring the current  ($I_{test}$) through the object. The impedance ($Z_{test}$) of the test object is then calculated using the equation ($Z_{test} = \frac{I_{known}}{I_{test}} * Z_{known}$).


\section {System Design: Gesture Detection Algorithm} \label{appendix:algorithm}
\subsection{System Design: Algorithm Parameters} \label{app:params}

\subsubsection{Data Augmentation: Equations of Noise}

\begin{enumerate}
    \item \emph{Shift}: $Z=Z\pm\delta$, which simulates the system bias caused by the individual difference and location bias of the electrodes. 
    \item \emph{Scale Up/Down}: $Z=\frac{Z*Z}{Z_{mean}}$ and $Z=\sqrt{Z}*\sqrt{Z_{mean}}$, which simulates the difference of bio-impedance variation caused by lung capacity and breathing habits. Here the $Z_{mean}$ is the average value of each experiment session.
    \item \emph{Gaussian Noise}: $Z=Z + N(\mu,\sigma^2)$, which simulates the random noise caused by the movement of electrodes caused by human activity.
\end{enumerate}

\subsubsection{Neural Network: Parameters} \label{app:nerualparams}
\begin{figure*}[!h]
\centering
  \begin{subfigure}[b]{0.40\textwidth}
  \includegraphics[width=1.0\linewidth]{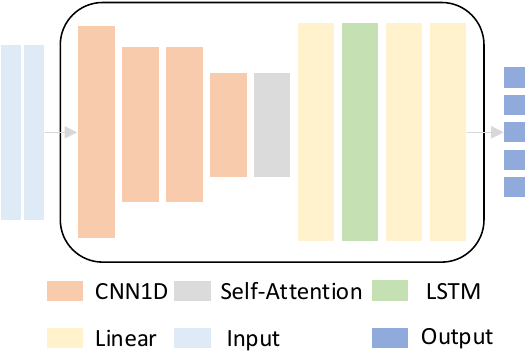} 
  \caption{BreathNet Architecture}
  \label{fig:model_arch}
  \end{subfigure}
  \begin{subtable}[b]{0.58\textwidth}
  \centering
    \begin{tabular}{c|c}
    \hline
    Layers & Parameters\\ 
    \hline

    \hline
    CNN1 & kernel\#: 100, kernel size: 3, stride: 1\\ \hline
    CNN2 & kernel\#: 100, kernel size: 3, stride: 2\\ \hline
    CNN3 & kernel\#: 100, kernel size: 3, stride: 1\\ \hline
    CNN4 & kernel\#: 100, kernel size: 3, stride: 2\\ \hline
    Self-Attention & embedding: 100, head \#: 20\\ \hline
    Linear1 & input:100, output:100\\ \hline
    LSTM & input:100, hidden\#:100\\ \hline
    Linear2 & input:2200, output:100\\ \hline
    Linear3 & input:100, output:5\\ \hline
    \end{tabular}
  \caption{BreathNet Parameters}
  \label{table:model_info}
  \end{subtable}
\caption{Model Information Summary}
\label{fig:model_summary}
\end{figure*}

\cref{fig:model_summary} presents a summary of the BreathNet model employed in the iBreath device for detecting breathing gestures. This model incorporates a sequence of one-dimensional convolutional (CNN1D), self-attention, long short-term memory (LSTM), and linear layers, as illustrated in \cref{fig:model_arch}. Detailed parameters of BreathNet are provided in \cref{table:model_info}. 

\subsubsection{Neural Network: Optimization Strategies of Time-Step Level Prediction Algorithm}

\begin{itemize}
    \item \emph{Low-pass Strategy}: If a non-null class prediction occurs only once and is flanked by null class predictions both before and after, it will be revised to a null class. Similarly, if a null class prediction is surrounded by non-null class predictions, it will be adjusted to match the preceding non-null class prediction. 
    \item \emph{Front follows back strategy}: If a single click gesture prediction is immediately succeeded by either a double click or SoS gesture prediction, the initial prediction will be updated to either double click or SoS gesture. Similarly, if a double click prediction is immediately followed by a triple click, the former prediction will be updated to triple click.
    \item \emph{Majority rule strategy}: In a sequence of consecutive non-null predictions, we use the principle of majority rule to correct the minority predictions to be the same as the majority predictions.
\end{itemize}

\insertappendix{System Design: Optimization Strategies of Time-Step Level Prediction Algorithm}{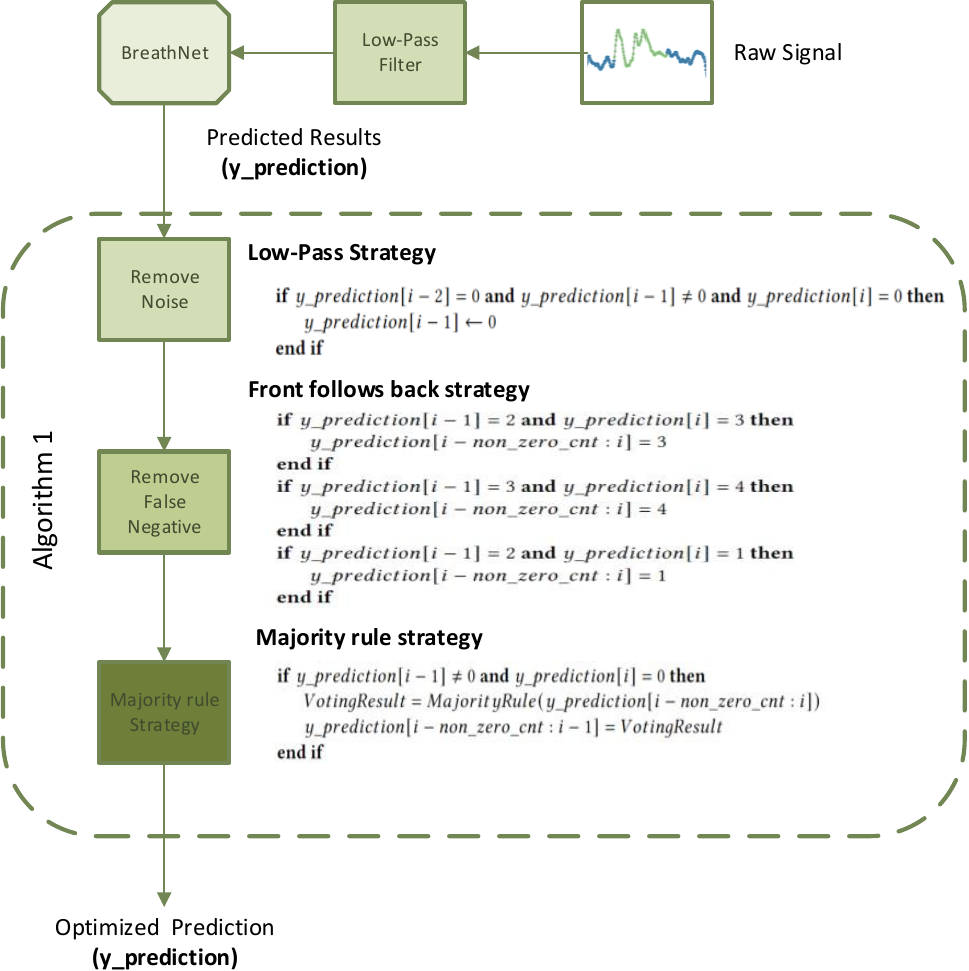}{fig: data_flow}
\subsection{System Design: Optimized Algorithm of Time-Step Level Prediction}
\label{sec: alorthm}
\begin{algorithm}

\caption{Optimization of Prediction Algorithm}
\begin{algorithmic}[1]
\footnotesize
\label{al:optimizing_prediction}
\State \textbf{Input:} $y\_prediction, strategy1, strategy2, strategy3$
\State $non\_zero\_cnt \gets 0$
\For{$i \gets 0$ \textbf{to} $len(y\_prediction) - 1$}
    \If{$i > 2$}
        
        \If{$strategy1$} \Comment{Execute strategy1: Low-Pass Strategy}
            \If{$y\_prediction[i-2] = 0$ \textbf{and} $y\_prediction[i-1] \neq 0$ \textbf{and} $y\_prediction[i] = 0$}
                \State $y\_prediction[i-1] \gets 0$
            \EndIf
        \EndIf
        \If{$y\_prediction[i-2] \neq 0$ \textbf{and} $y\_prediction[i-1] = 0$ \textbf{and} $y\_prediction[i] \neq 0$}
            \State $y\_prediction[i-1] \gets y\_prediction[i]$
        \EndIf
        
        \If{$strategy2$} \Comment{Execute strategy2: Front follows back strategy} 
            \If{$y\_prediction[i-1] = 2$ \textbf{and} $y\_prediction[i] = 3$}
                \State $y\_prediction[i-non\_zero\_cnt:i] =3$
            \EndIf

            \If{$y\_prediction[i-1] = 3$ \textbf{and} $y\_prediction[i] = 4$}
                \State $y\_prediction[i-non\_zero\_cnt:i] =4$
            \EndIf

            \If{$y\_prediction[i-1] = 2$ \textbf{and} $y\_prediction[i] = 1$}
                \State $y\_prediction[i-non\_zero\_cnt:i] =1$
            \EndIf
        \EndIf



        \If{$strategy3$} \Comment{Execute strategy3: Majority rule strategy}
            \If{$y\_prediction[i-1] \neq 0$ \textbf{and} $y\_prediction[i] = 0$}
                \State $VotingResult = MajorityRule(y\_prediction[i-non\_zero\_cnt:i])$ 
                \State $y\_prediction[i-non\_zero\_cnt:i-1] = VotingResult$ 
            \EndIf
        \EndIf

        \If{$y\_prediction[i-1] \neq 0$ \textbf{and} $y\_prediction[i] = 0$}

            \State $non\_zero\_cnt \gets 0$
        \Else
            \State $non\_zero\_cnt \gets non\_zero\_cnt + 1$
        \EndIf

    \EndIf
\EndFor

\end{algorithmic}
\end{algorithm}

\section {Part 2: UX Study Material}
\subsection{Interview Script}
\label{sec: Interview}

\begin{enumerate}[itemsep=9mm]
    \item \textbf{Perceived effort}:
        \begin{enumerate}
            \item How are you feeling? Can you describe your overall experience with the breathing patterns?
        \end{enumerate}

    \item \textbf{Learnability}:
        \begin{enumerate}
            \item How easy or hard was it for you to understand the graph and breathe following the ball? (Follow up: why? / explain more)
            \item How easy or difficult was it for you to learn the patterns and do them alone without the graph? (Follow up: were you able to remember the patterns when there was no graph?)
            \item Do you think you would be able to learn a new breathing pattern just by looking at a new graph without explanation? (if they said no: what would you change?)
            \item Did you feel you need less or more time to train before doing the breathing alone without the graph? (follow up: how many times would be optimal for you?)
        \end{enumerate}

    \item \textbf{Complexity}:
        \begin{enumerate}
            \item Were there aspects that you found particularly challenging during the experiment?
            \item Were there aspects that you found particularly intuitive during the experiment?
        \end{enumerate}
        
    \item \textbf{Application areas}:
        \begin{enumerate}
            \item Would you consider using the breathing pattern as a way to click on a button? (explain to them by acting it if they don’t understand) (follow up: why?)
            \item In which scenarios would you want to click a button using the breathing techniques?
            \item Can you provide examples of specific tasks where the breathing patterns can be used to interact with systems or your environment?
            \item On a scale from 1 to 10, how satisfied are you with the new interaction technique, and why?
        \end{enumerate}

    \item \textbf{Improvements}:
        \begin{enumerate}
            \item How well do you think the new interaction technique can be integrated into your regular workflow or daily activities?
            \item If you could change or improve one thing about the new interaction technique, what would it be? (if they don’t understand the interaction technique, tell them the using breathing to interact with your environment
            \item Do you foresee any challenges or barriers to adopting this new technique in your work or personal life?
        \end{enumerate}
        
\end{enumerate}
\insertsurvey{Questionnaire}{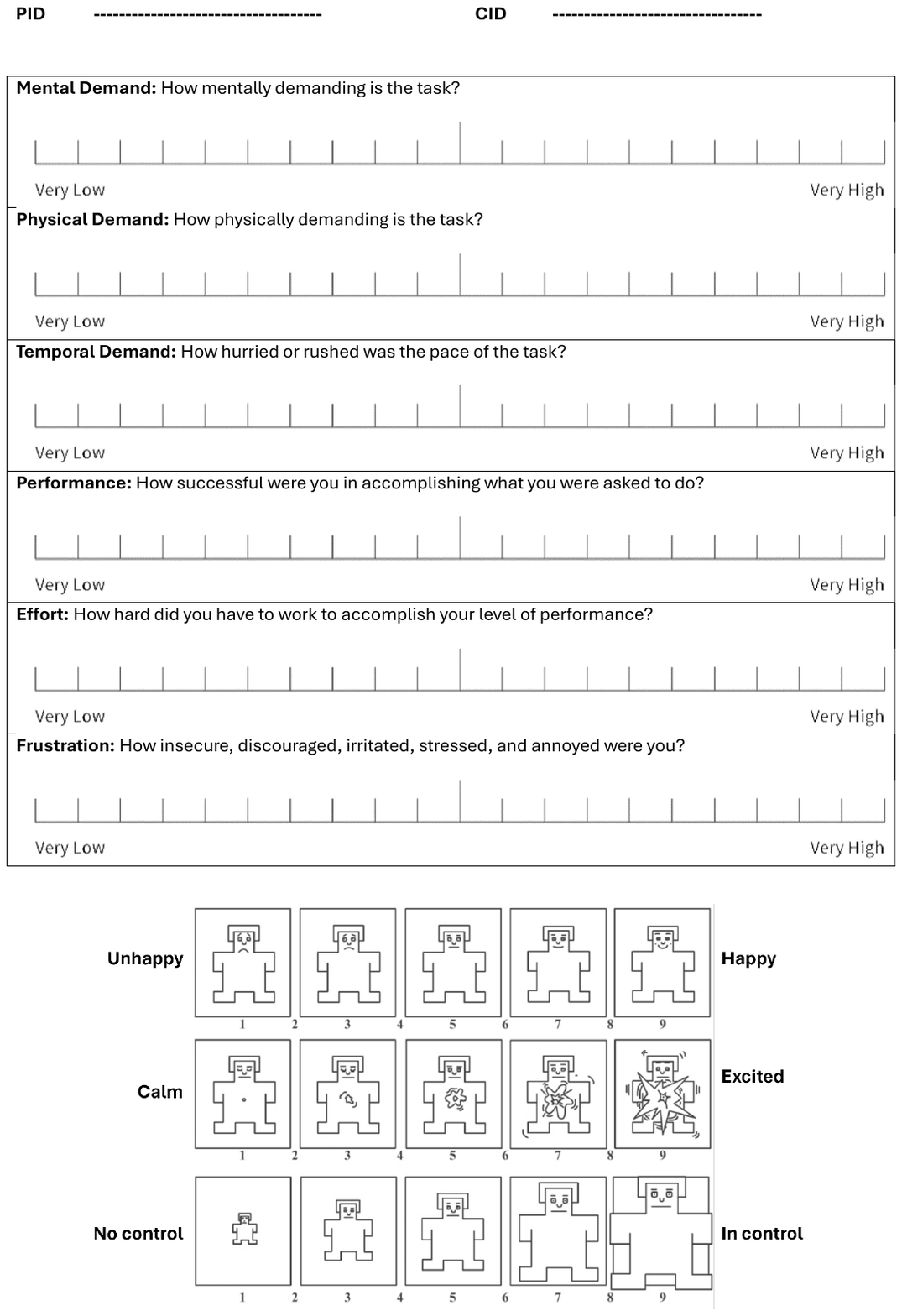}{sec: Questionnaire}










\end{document}